\documentclass{article}
\usepackage{amssymb}
\newtheorem{thm}{Theorem}[section]
\newtheorem{prop}[thm]{Proposition}
\newtheorem{lem}[thm]{Lemma}

\newcommand{\comment}[1]{}
\newcommand{\BZ}{{\Bbb Z}}

\newcommand{\BC}{{\Bbb C}}

\def\qed{\rule{1mm}{2.5mm}}

\newcommand{\IV}{\mbox{\scriptsize IV}}

\newcommand{\dfrac}[2]{{\displaystyle\frac{#1}{#2}}}
\newcommand{\br}[1]{\langle #1 \rangle}
\title{
{\bf A study on the fourth $q$-Painlev\'e equation}
}
\author{
{\normalsize
Kenji KAJIWARA${}^{1)}$, 
Masatoshi NOUMI${}^{2)}$ and Yasuhiko YAMADA${^{2)}}$
}\cr
{\small 
${}^{1)}$ Department of Electrical Engineering, Doshisha University}\cr
{\small Kyotanabe, Kyoto 610-0321, JAPAN}\cr
{\small  ${}^{2)}$ Department of Mathematics, Kobe University}\cr
{\small Rokko, Kobe 657-8501, JAPAN}\cr }
\begin{document}
\maketitle
\begin{abstract}
 A $q$-difference analogue of the fourth Painlev\'e equation is
 proposed. Its symmetry structure and some particular solutions are investigated.
\end{abstract}

\section*{Introduction}
The importance of six Painlev\'e equations and their solutions in
mathematics and mathematical physics is widely accepted.  The discrete
analogue of Painlev\'e equations has been first recognized in \cite{JMU,JM1,JM2}
where they appear as Schlesinger transformations of the Painlev\'e
equations. After the discovery of discrete analogue of the Painlev\'e
property which is now called the singularity confinement
property\cite{GRP}, a lot of second order difference equations,
including $q$-difference equations, have been identified as the discrete
Painlev\'e equations\cite{RGH,GR:physicaA}. By investigating the Lax
pairs, particular solutions including $\tau$ functions and bilinear
equations, and so on\cite{GNR:century}, it has been gradually recognized
that discrete Painlev\'e equations admit similar properties to original
Painlev\'e equations.

The purpose of this paper is to introduce a $q$-difference analogue of
the fourth Painlev\'e equation P$_{\rm IV}$. This equation, called the
$q$-Painlev\'e IV equation $q$P$_{\rm IV}$ below, shares many
characteristic properties with the original P$_{\rm IV}$. In particular,
it admits the action of the affine Weyl group of type $A^{(1)}_2$ as a
group of B\"acklund transformations. Furthermore, it has an analogue of
classical solutions expressible by the continuous $q$-Hermite-Weber
functions and rational solutions corresponding those of P$_{\rm IV}$
studied in \cite{Okamoto:P4,KO,NY2}. 

The plan of this paper is as follows. We introduce our $q$P$_{\rm IV}$
in Section 1 and describe its symmetry structure in terms of the affine
Weyl group $W(A^{(1)}_2)$. In Section 2, we construct a $q$-analogue of
classical solutions along each reflection hyperplane from a seed
solution which is described by a discrete Riccati equation. Each
particular solutions of this class is expressed in terms of a Toeplitz
type determinant of continuous $q$-Hermite-Weber functions. 
In Section 3, we discuss some related topics including the relationship 
between our $q$P$_{\rm IV}$ and Sakai's Mul.6 system \cite{Sak} and the 
ultra-discrete limit of $q$P$_{\IV}$.

%%%%%%%%%%%%%%%%%%%%%%%%%%%%%%%%%%%%%%%%%%%%%%%%%%
%%%   section 1
%%%%%%%%%%%%%%%%%%%%%%%%%%%%%%%%%%%%%%%%%%%%%%%%%%

\section{Symmetric form of a $q$-Painlev\'e equation $q$P$_{\IV}$}

\subsection{$q$-Painlev\'e equation $q$P$_{\IV}$}

In this paper, we consider the following discrete system 
including  three dependent variables $f_0$, $f_1$, $f_2$ and 
three parameters $a_0$, $a_1$, $a_2$: 
\begin{equation}\label{qP4}
\begin{array}{l}
\smallskip
\overline{a_0}=a_0,\quad
\overline{a_1}=a_1,\quad
\overline{a_2}=a_2,\quad\cr
\smallskip
\overline{f_0}=a_0a_1f_1
\dfrac{1+a_2f_2+a_2a_0f_2f_0}{
1+a_0f_0+a_0a_1f_0f_1},\cr
\smallskip
\overline{f_1}=a_1a_2f_2
\dfrac{1+a_0f_0+a_0a_1f_0f_1}{
1+a_1f_1+a_1a_2f_1f_2},\cr
\overline{f_2}=a_2a_0f_0
\dfrac{1+a_1f_1+a_1a_2f_1f_2}{
1+a_2f_2+a_2a_0f_2f_0},
\end{array}
\end{equation}
where $\overline{\phantom{a}}$ stands for the discrete time evolution. 
We will also use the notation $t(x)=\overline{x}$ 
when we regard $x\to \overline{x}$ as a transformation of variables. 
The inverse transformation of $t$ is given by
\begin{equation}\label{qP4:inverse}
\begin{array}{l}
\smallskip
\underline{a_0}=a_0,\quad
\underline{a_1}=a_1,\quad
\underline{a_2}=a_2,\quad\cr
\smallskip
\underline{f_0}=\dfrac{f_2}{a_0a_1}
\dfrac{a_0a_1+a_0f_1+f_0f_1}{a_2a_0+a_2f_0+f_2f_0},\cr
\smallskip
\underline{f_1}=\dfrac{f_0}{a_1a_2}
\dfrac{a_1a_2+a_1f_2+f_1f_2}{a_0a_1+a_0f_1+f_0f_1},\cr
\underline{f_2}=\dfrac{f_1}{a_2a_0}
\dfrac{a_2a_0+a_2f_0+f_2f_0}{a_1a_2+a_1f_2+f_1f_2}.
\end{array}
\end{equation}
%We introduce a constant $q$ by setting $a_0a_1a_2=q$. 
%Noting that the product $f_0f_1f_2$ can be regarded as the independent 
%variable, we introduce a variable $c$ such that $f_0f_1f_2=q c^2$ and $\overline{c}=q c$.  
%If we regard $f_j$ ($j=0,1,2$) as functions in $c$, eq.(\ref{qP4}) 
%thus represents a system of $q$-difference equations for the unknown 
%functions $f_j=f_j(c)$ ($j=0,1,2$) with parameters $a_j$ ($j=0,1,2$) 
%such that $a_0a_1a_2=q$.  

We introduce a constant $q$ by setting $a_0a_1a_2=q$. 
Noting that the product $f_0f_1f_2$ can be regarded as the independent 
variable, we introduce a variable $c$ such that $f_0f_1f_2=q c^2$ and $\overline{c}=q c$.  
If we regard $f_j$ ($j=0,1,2$) as functions in $c$, eq.(\ref{qP4}) 
thus represents a system of $q$-difference equations for the unknown 
functions $f_j=f_j(c)$ ($j=0,1,2$) with parameters $a_j$ ($j=0,1,2$) 
such that $a_0a_1a_2=q$. 
As we will see in Subsection 1.3 below, 
this $q$-difference system has a continuous limit to the 
{\em symmetric form} of the fourth 
Painlev\'e equation $P_{\IV}$.   
For this reason, we refer to eq.(\ref{qP4}) as the symmetric 
form of the fourth $q$-Painlev\'e equation $q$P$_{\IV}$. 

%\subsection{B\"acklund transformations: $W(A^{(1)}_2)$}
\subsection{B\"acklund transformations}
The discrete system eq.(\ref{qP4}) admits the action of the (extended)
affine Weyl group $\widetilde{W}=\br{s_0,s_1,s_2,\pi}$ of 
type $A^{(1)}_2$ as a group of B\"acklund transformations. 
In what follows we denote by $\widetilde{W}$ the group 
generate by the generators $s_0,s_1,s_2,\pi$ and the 
the fundamental relations
\begin{equation}\label{affWeyl}
s_i^2=1,\quad
(s_is_{i+1})^3=1,\quad
\pi^3=1,\quad
\pi s_i =s_{i+1}\pi\quad(i=0,1,2),
\end{equation}
where the indices are understood as elements 
of $\BZ/3\BZ$.  

We define the action of 
$s_0$, $s_1$, $s_2$, 
$\pi$ on the parameters $a_0$, $a_1$, $a_2$ 
by the following formulas:
\begin{equation}\label{Wona}
\begin{array}{lll}
s_0(a_0)=a_0^{-1}, & s_0(a_1)=a_1a_0, & s_0(a_2)=a_2a_0,\cr
s_1(a_0)=a_0a_1, & s_1(a_1)=a_1^{-1},& s_1(a_2)=a_2a_1,\cr
s_2(a_0)=a_0a_2, & s_2(a_1)=a_1a_2, & s_2(a_2)=a_2^{-1},\cr
\ \pi(a_0)=a_1, &\ \pi(a_1)=a_2, & \ \pi(a_2)=a_0. 
\end{array}
\end{equation}
Note that this definition is a multiplicative version of 
the standard realization of 
$\widetilde{W}$ on the {\em simple roots} $\alpha_j$
defined through $a_j=q^{\alpha_j}$ ($j=0,1,2$). 
We now define the action of 
$s_0$, $s_1$, $s_2$, 
$\pi$ on the $f$ variables as follows:
\begin{equation}\label{Wonf}
\begin{array}{lll}
\smallskip
s_0(f_0)=f_0 ,
& s_0(f_1)=f_1\,\dfrac{a_0+f_0}{1+a_0f_0} ,
& s_0(f_2)=f_2\,\dfrac{1+a_0f_0}{a_0+f_0},\cr
\smallskip
s_1(f_0)=f_0\,\dfrac{1+a_1f_1}{a_1+f_1},
& s_1(f_1)=f_1, & 
s_1(f_2)=f_2\,\dfrac{a_1+f_1}{1+a_1f_1},\cr
\smallskip
s_2(f_0)=f_0\,\dfrac{a_2+f_2}{1+a_2f_2},& 
s_2(f_1)=f_1\,\dfrac{1+a_2f_2}{a_2+f_2},& s_2(f_2)=f_2,\cr
\ \pi(f_0)=f_1, &\ \pi(f_1)=f_2, & \ \pi(f_2)=f_0. 
\end{array}
\end{equation}
We remark that the action of $s_i$ on the $a$-variables and
the $f$-variables is given by 
\begin{equation} \label{W:general}
s_i(a_j)=a_j a_i^{-a_{ij}},\quad
s_i(f_j)=f_j\,\left(\frac{a_i+f_i}{1+a_if_i}\right)^{u_{ij}}
\qquad(i,j=0,1,2),
\end{equation}
respectively, 
where $A=(a_{ij})_{i,j=0}^2$ is the generalized Cartan matrix 
of type $A^{(1)}_2$
and $U=(u_{ij})_{i,j=0}^2$ is an orientation matrix of the 
corresponding Dynkin diagram:
\begin{equation}\label{AandU}
A=\left[\begin{array}{ccc}
2 & -1 & -1 \cr
-1 & 2 & -1 \cr
-1 & -1 &  2 
\end{array}\right],\quad
U=\left[\begin{array}{ccc}
0 & 1 & -1 \cr
-1 & 0 & 1 \cr
1 & -1 &  0
\end{array}\right]. 
\end{equation}
The following theorem can be verified by direct computation. 
\begin{thm}
The transformations $s_0$, $s_1$, $s_2$ and $\pi$ 
of the $a$-variables and the $f$-variables, 
defined by $(\ref{Wona})$, $(\ref{Wonf})$, 
generate the extended affine Weyl group 
$\widetilde{W}=\br{s_0,s_1,s_2,\pi}$ of type $A^{(1)}_2$. 
Furthermore, they commute with the time evolution $t$ 
of the fourth $q$-Painlev\'e equation $q$P$_{\IV}$. 
\end{thm}

By using the action of the extended affine Weyl group $\widetilde{W}$,
we can define the {\em Schlesinger transformations\,} $T_1$, $T_2$, 
$T_3$ for $q$P$_{\IV}$ as 
\begin{equation}
T_1=\pi s_2 s_1,\quad T_2=s_1 \pi s_2,\quad T_3=s_2 s_1 \pi. 
\end{equation}
Note that 
$T_iT_j=T_jT_i$ ($i,j=1,2,3$)  
and $T_1T_2T_3=1$. 
The action of $T_1$ on the variables $a_j$, $f_j$ is given explicitly 
as follows:
\begin{equation}\label{T1}
\begin{array}{l}
\smallskip
T_1(a_0)=q a_0 ,\quad T_1(a_1)=q^{-1}a_1,\quad T_1(a_2)=a_2,\cr
\smallskip
T_1(f_0)=f_1\,\dfrac
{(a_0+f_0)(a_0+f_0+a_0a_2 f_2+a_0^2 a_2 f_0 f_2)}
{(1+a_0f_0)(a_0^2a_2+a_0a_2 f_0+f_2+a_0 f_0 f_2)},
\cr
\smallskip
T_1(f_1)=f_2\,\dfrac{1+a_0f_0}{a_0+f_0},
\cr
T_1(f_2)=f_0\,\dfrac
{a_0^2a_2+a_0a_2 f_0+f_2+a_0 f_0 f_2}
{a_0+f_0+a_0a_2 f_2+a_0^2 a_2 f_0 f_2}. 
\end{array}
\end{equation}
The corresponding formulas for $T_2$ and $T_3$ are obtained 
by the rotation of indices, since 
$T_2=\pi T_1\pi^{-1}$ and $T_3=\pi T_2\pi^{-1}$. 
Each of these Schlesinger transformations commutes with the 
time evolution of $q$P$_{\IV}$ and can be regarded as a version of
the third $q$-Painlev\'e equation. For example, eq.(\ref{T1}) is 
rewritten equivalently as
\begin{equation}\label{qP3}
 \begin{array}{l}
\smallskip
 T_1(f_1)=f_2~\dfrac{1+a_0f_0}{a_0+f_0}=\dfrac{qc^2}{f_0f_1}~\dfrac{1+a_0f_0}{a_0+f_0},\\
 T_1^{-1}(f_0)=f_2~\dfrac{a_1+f_1}{1+a_1f_1}=\dfrac{qc^2}{f_0f_1}~\dfrac{a_1+f_1}{1+a_1f_1}. 
 \end{array}
\end{equation}
This is an analogue of the fact that some discrete Painlev\'e equations
arise from B\"acklund transformations of the Painlev\'e equations\cite{GNR:century,JMU,JM1,JM2,NY1}.
\subsection{Limit transition to $P_{\IV}$}
By introducing a small parameter $\varepsilon$ such that
$q=e^{-\varepsilon^2/2}$, we set
\begin{equation}
a_i=e^{-\varepsilon^2 \alpha_i/2},\quad
f_i=-e^{-\varepsilon \varphi_i}\qquad (j=0,1,2).
\end{equation}
Then we have 
\begin{equation}
\overline{f_i}-f_i=
\varepsilon^2 
\big(\varphi_i (\varphi_{i+1}-\varphi_{i+2})+\alpha_i\big)+O(\varepsilon^3)
\end{equation}
for $i=0,1,2$. 
Passing to the variables $\alpha_j$ and $\varphi_j$, we define 
the derivation $'$ by
\begin{equation}
x'=\lim_{\varepsilon\to 0}\frac{1}{\varepsilon} (\overline{x}-x)
\end{equation}
for a function $x$ in $\alpha_j$ and $\varphi_j$. 
Then we have 
$\overline{f_i}-f_i=\varepsilon^2\varphi_i'+O(\varepsilon^3)$. 
Hence we obtain 
\begin{equation}\label{sP4}
\begin{array}{l}
\smallskip
\alpha_0'=0,\quad \alpha_1'=0,\quad \alpha_2'=0,\cr
\smallskip
\varphi_0'=\varphi_0(\varphi_1-\varphi_2)+\alpha_0, \nonumber\cr
\smallskip
\varphi_1'=\varphi_1(\varphi_2-\varphi_0)+\alpha_1, \cr
\varphi_2'=\varphi_2(\varphi_3-\varphi_1)+\alpha_2. 
\end{array}
\end{equation}
This differential system is the symmetric form the 
fourth Painlev\'e equation $P_{\IV}$
\cite{VS,A,NY3}.  
In fact, under the normalization 
$\varphi_0+\varphi_1+\varphi_2=t$, $\alpha_0+\alpha_1+\alpha_2=1$, 
the system (\ref{sP4}) is equivalent to the second order differential equation 
\begin{equation}
y''=\frac{1}{2y}\big(y'\big)^2
+\frac{3}{2}y^3-2 t y^2+\big(\frac{t^2}{2}+\alpha_1-\alpha_2\big)y
-\frac{\alpha_0^2}{2y}
\end{equation}
for $y=\varphi_0$, where $'=d/dt$. 
Through the limiting procedure, the B\"acklund transformations 
$s_0$, $s_1$, $s_2$, $\pi$ for our $q$P$_{\IV}$ 
also pass to those for the symmetric 
form of $P_{\IV}$ (\ref{sP4}) such 
that
\begin{equation}
\begin{array}{ll}
s_i(\alpha_j)=\alpha_j-\alpha_i a_{ij},&
s_i(\varphi_j)=\varphi_j+\dfrac{\alpha_i}{\varphi_i} u_{ij}, \\
\pi(\alpha_j)=\alpha_{j+1},&\pi(\varphi_j)=\varphi_{j+1},
\end{array}
\quad (i,j=0,1,2)
\end{equation}
as in \cite{NY2, NY3}. 

%%%%%%%%%%%%%%%%%%%%%%%%%%%%%%%%%%%%%%%%%%%%%%%%%%
%%%   section 2
%%%%%%%%%%%%%%%%%%%%%%%%%%%%%%%%%%%%%%%%%%%%%%%%%%
%\section{Classical solutions of $q$P$_{\IV}$}
\section{Particular solutions for $q$P$_{\IV}$}
In this section, we investigate $q$-analogue of classical solutions of P$_{\IV}$ and 
determinant formulas for them. 
Recall that P$_{\IV}$ has two classes of classical solutions \cite{Okamoto:P4,KO,NY3},
those of hypergeometric type, expressed in terms of 
Hermite-Weber functions and the rational solutions. 

\subsection{Continuous $q$-Hermite-Weber functions as seed solutions}
% In this section, we construct particular solutions to $q$P$_{\IV}$.  
In order to show the parameter dependence explicitly, we rewrite $q$P$_{\IV}$
(\ref{qP4}) by operating $T_1^\nu T_2^N$ on (\ref{qP4})
($\nu,N\in\BZ$). We denote $T_1^\nu T_2^N(f_i)=f_i(c;\nu,N)$, and we
abbreviate the unnecessary arguments depending on the context.  
Notice that
\begin{equation}
T_1^\nu T_2^N(a_0) = a_0q^{\nu},\quad T_1^\nu T_2^N(a_1) = a_1q^{-\nu+N},\quad 
T_1^\nu T_2^N(a_2) = a_2q^{-N},\quad T_1^\nu T_2^N(c) = c,
\end{equation}
and
\begin{equation}
 t(a_i)=a_i,\ (i=0,1,2),\quad t(c)=qc,
\end{equation}
where 
\begin{equation}
q=a_0a_1a_2,\quad c=\left(\frac{f_0f_1f_2}{q}\right)^{1/2}.\label{normalization}
\end{equation}
The $q$-Painlev\'e IV equation $q$P$_{\IV}$ with respect to the variable $c$ is expressed as
\begin{eqnarray}
&&f_0(qc)=a_0a_1q^{N}f_1(c)
\dfrac{1+a_2q^{-N}f_2(c)+a_2a_0q^{\nu-N}f_2(c)f_0(c)}{
1+a_0q^\nu f_0(c)+a_0a_1q^{N}f_0(c)f_1(c)},\label{qP4:1}\\
&&f_1(qc)=a_1a_2q^{-\nu}f_2(c)
\dfrac{1+a_0q^\nu f_0(c)+a_0a_1q^{N}f_0(c)f_1(c)}{
1+a_1q^{-\nu+N}f_1(c)+a_1a_2q^{-\nu}f_1(c)f_2(c)},\label{qP4:2}\\
&&f_2(qc)=a_2a_0q^{\nu-N}f_0(c)
\dfrac{1+a_1q^{-\nu+N}f_1(c)+a_1a_2q^{-\nu}f_1(c)f_2(c)}{
1+a_2q^{-N}f_2(c)+a_2a_0q^{\nu-N}f_2(c)f_0(c)}. \label{qP4:3}
\end{eqnarray}
where
the arguments $\nu$, $N$ for $f_i$ are suppressed so that  $f_i(c)=f_i(c;\nu,N)$.
In what follows, we use similar abbreviations 
\begin{equation}
f_i(q^k c)=f_i(q^kc;\nu,N),\quad
f_i(\nu+k)=f_i(c;\nu+k,N),\quad
f_i(N+k)=f_i(c;\nu,N+k)
\end{equation}
for $k\in\BZ$. 

First let us consider the case $N=0$. It is possible to specialize the variables as
\begin{equation}
 f_2=-1,\quad a_2=1,\label{specialization}
\end{equation}
consistently. In fact, eq.(\ref{qP4:3}) becomes trivial by this
specialization, and eqs.(\ref{qP4:1}) and (\ref{qP4:2}) are reduced to a
discrete Riccati type equation
\begin{equation}
 f_1(qc)=-\frac{q}{a_0q^{\nu}}\frac{(1-q^2c^2)f_1(c)-a_0q^{\nu}c^2}{f_1(c)}
\label{dRiccati:c}
\end{equation}
and $f_0(c)=-\frac{qc^2}{f_1(c)}$. By putting
$f_1(c)=\frac{F_\nu(c)}{G_\nu(c)}$ in eq.(\ref{dRiccati:c}), we can
solve eq.(\ref{dRiccati:c}) as
\begin{equation}
  f_1(c) = -\frac{q}{a_0q^{\nu}}\frac{G_\nu(qc)}{G_\nu(c)},\label{dep0:f1}
\end{equation}
with $G_\nu(c)$ satisfying the linear $q$-difference equation
\begin{equation}
G_\nu(q^2c)=(1-q^2c^2)G_\nu(qc)+a_0^2q^{2\nu}c^2 G_\nu(c).\label{contiguity:c}
\end{equation}

Let us derive contiguity relations to be satisfied by $G_\nu(c)$ in the direction of
$\nu$. For this purpose, we consider the Schlesinger
transformation $T_1$ (\ref{T1}),
\begin{eqnarray}
f_0(\nu+1)&=&f_1(\nu)\,\dfrac{a_0q^{\nu}+f_0(\nu)}{1+a_0q^{\nu}f_0(\nu)}\nonumber\\
&&\times\dfrac
{a_0q^{\nu}+f_0(\nu)+a_0a_2q^{\nu-N} f_2(\nu)+a_0^2 a_2q^{2\nu-N} f_0(\nu)
 f_2(\nu)}
{a_0^2a_2q^{2\nu-N}+a_0a_2q^{\nu-N} f_0(\nu)+f_2(\nu)+a_0q^{\nu} 
f_0(\nu) f_2(\nu)},\label{T1:1}\\
f_1(\nu+1)&=&f_2(\nu)\,\dfrac{1+a_0q^{\nu}f_0(\nu)}{a_0q^\nu+f_0(\nu)},\label{T1:2}\\
f_2(\nu+1)&=&f_0(\nu)\nonumber\\
&&\times\dfrac
{a_0^2a_2q^{2\nu-N}+a_0a_2q^{\nu-N} f_0(\nu)+f_2(\nu)+a_0q^\nu f_0(\nu) f_2(\nu)}
{a_0q^\nu+f_0(\nu)+a_0a_2q^{\nu-N} f_2(\nu)+a_0^2 a_2q^{2\nu-N} f_0(\nu) f_2(\nu)}.\label{T1:3}
\end{eqnarray}
Applying the specialization (\ref{specialization}) and putting $N=0$,
eqs.(\ref{T1:1})-(\ref{T1:3}) are reduced to a discrete Riccati type equation,
\begin{equation}
 f_1(\nu+1)=-\,\dfrac{f_1(\nu) -a_0q^{\nu+1}c^2}{a_0q^\nu f_1(\nu) - qc^2}.\label{dRiccati:nu}
\end{equation}
Substituting eq.(\ref{dep0:f1}) into eq.(\ref{dRiccati:nu}), we obtain
contiguity relations for $G_\nu(c)$,
\begin{eqnarray}
 &&G_{\nu+1}(qc)=G_{\nu}(qc)+a_0^2q^{2\nu}c^2G_\nu(c),\label{contiguity:cn1}\\
 &&G_{\nu+1}(c)=G_{\nu}(qc)+c^2G_\nu(c).\label{contiguity:cn2}
\end{eqnarray}
In particular, we obtain a three-term relation in the direction of $\nu$,
\begin{equation}
G_{\nu+2}(c) = \left(c^2+1\right)G_{\nu+1}(c) - c^2(1-a_0^2q^{2\nu})G_\nu(c).\label{contiguity:nu}
\end{equation}
We note that eqs. (\ref{contiguity:c}) and (\ref{contiguity:nu}) are
derived from eqs.(\ref{contiguity:cn1}) and (\ref{contiguity:cn2}).

{}From the above discussion, we have:
\begin{prop}\label{seed}
 $q$P$_{\IV}$ (\ref{qP4:1})-(\ref{qP4:3}) admits 1-parameter family of
 particular solutions given by
\begin{equation}
f_0 =c^2a_0q^{\nu}\frac{G_\nu(c)}{G_\nu(qc)},\quad 
f_1 = -\frac{q}{a_0q^{\nu}}\frac{G_\nu(qc)}{G_\nu(c)},\quad f_2=-1
\end{equation}
for $N=0$, where $G_\nu(c)$ is a function 
satisfying the contiguity relations (\ref{contiguity:cn1}) and (\ref{contiguity:cn2}).
\end{prop}
It is interesting to note that eq.(\ref{contiguity:c}) admits polynomial
solutions in $c$ if $a_0=q$,
\begin{eqnarray*}
G_0(c)&=&1,\\
G_1(c)&=&c^2+1,\\
G_2(c)&=&c^4+(1+q^2)c^2+1,\\
G_3(c)&=&c^6+(1+q^2+q^4)c^4+(1+q^2+q^4)c^2+1,\\
&\cdots&\ .
\end{eqnarray*}
It is not difficult to check that 
\begin{equation}
G_\nu(c)=\sum_{k=0}^\nu\frac{(q^2;q^2)_\nu}{(q^2;q^2)_k(q^2;q^2)_{\nu-k}}c^{2\nu-2k}
=\sum_{k=0}^\nu\left[{\nu\atop k}\right]_{q^2}c^{2\nu-2k}\quad (\nu=0,1,2,\cdots)
\end{equation}
satisfy eqs. (\ref{contiguity:cn1}) and (\ref{contiguity:cn2}), where
$\left[{\nu\atop k}\right]_q$ is the $q$-binomial coefficient. The generating function for
these polynomials is given by
\begin{equation}
 \frac{1}{(c\lambda;q^2)(\lambda/c;q^2)}=\sum_{\nu=0}^\infty \frac{G_\nu(c)}{(q^2;q^2)_\nu}
\left(\frac{\lambda}{c}\right)^\nu.
\end{equation}
The polynomials $H_\nu(x)=c^{-\nu}G_\nu(c)$, $x=\frac{c+c^{-1}}{2}$ are
called {\em the continuous $q$-Hermite polynomials} \cite{KS,GR}. In this sense,
$c^{-\nu}G_\nu(c)$ for $a_0\neq q^l$ $(l\in \BZ)$ may be regarded as a
$q$-difference analogue of the Hermite-Weber functions,  to which hereafter we will
refer as ``continuous $q$-Hermite-Weber functions''.
%\par\medskip

We note that it is possible to obtain ``higher order'' solutions for
$N\in\BZ$ by successive applications of $T_2$ on the solutions obtained
above\footnote{The situation is a little delicate for $N<0$, but we
will discuss in Section 2.2.3.}. These solutions are expressed rationally in the
continuous $q$-Hermite-Weber functions. 

{}From the rotational symmetry $\pi$ of $q$P$_{\IV}$, it is possible to
apply other specialization, $(a_0,f_0)=(1,-1)$, $\nu=0$ or
$(a_1,f_1)=(1,-1)$, $\nu-N=0$ on $q$P$_{\IV}$(\ref{qP4:1})-(\ref{qP4:3})
and perform the same procedure as discussed above. Therefore, we obtain
\begin{thm}
When $a_i=q^k$ for some $i=0,1,2$ and $k\in\BZ$, the 
fourth $q$-Painlev\'e equation 
$q$P$_{\IV}$ (\ref{qP4}) admits a 1-parameter family of particular solutions
 which are expressed rationally by the continuous
 $q$-Hermite-Weber functions. 
\end{thm}
Explicit description of these solutions will be given in the next section.

Note that the hypersurfaces $a_i=q^k$ ($i=0,1,2$; $k\in\BZ$) in the parameter space 
correspond to the reflection hyperplanes of the affine Weyl group $W=W(A^{(1)}_2)$. 
This is an analogue of a well-known result by Okamoto for the classical solutions of the 
continuous P$_{\IV}$\cite{Okamoto:P4}.
\subsection{Toeplitz determinants}
\subsubsection{Determinant formula and bilinear equations}
In this section, we prove the following determinant formula for the
particular solutions obtained in the previous section.
\begin{thm}\label{Toeplitz}
 Let $G_\nu(c)$ is a solution of
eqs. (\ref{contiguity:cn1}) and (\ref{contiguity:cn2}).  
For each $N\in\BZ_{\ge 0}$, 
we define an $N\times N$
Toeplitz determinant 
\begin{equation}
 \phi_N^\nu(c)=
\det \left(G_{\nu-i+j}(c)\right)_{i,j=1,\cdots,N}
\quad (N\in\BZ_{>0}),\quad  \phi_0^\nu(c)=1.
\end{equation}
Then,
\begin{eqnarray}
&& f_0(c;\nu,N)=a_0q^\nu c^2~\frac{\phi_{N+1}^\nu(c)\phi_N^\nu(qc)}{\phi_{N+1}^\nu(qc)\phi_N^\nu(c)},
\label{f0}\\
&& f_1(c;\nu,N)= -\frac{1}{a_0q^{\nu+N-1}}~\frac{\phi_{N+1}^\nu(qc)\phi_N^{\nu-1}(c)}{\phi_{N+1}^\nu(c)\phi_N^{\nu-1}(qc)},\label{f1}\\
&& f_2(c;\nu,N)=-q^N~\frac{\phi_{N}^{\nu-1}(qc)\phi_N^\nu(c)}{\phi_{N}^\nu(qc)\phi_N^{\nu-1}(c)}\label{f2}
\end{eqnarray}
satisfy $q$P$_{\IV}$(\ref{qP4:1})-(\ref{qP4:3}) with $a_2=1$.
\end{thm}
We note that the case of $N=0$ agrees with Proposition \ref{seed}.

Theorem \ref{Toeplitz} is a direct consequence of the following
proposition:
\begin{prop}\label{bilinear}
 $\phi_N^\nu(c)$ satisfies the following bilinear difference equations,
\begin{eqnarray}
&& a_0^2c^2q^{2\nu}\phi_N^\nu(qc)\phi_{N+1}^\nu(c) + \phi_N^\nu(c)\phi_{N+1}^\nu(qc)
=q^{-2N}\phi_{N+1}^{\nu+1}(qc)\phi_N^{\nu-1}(c),\label{bl1}\\
&& a_0^2q^{2\nu}\phi_N^{\nu}(qc)\phi_{N+1}^{\nu+1}(c) -\phi_N^{\nu}(c)\phi_{N+1}^{\nu+1}(qc)
=(a_0^2q^{2\nu}-q^{2N})\phi_{N+1}^{\nu}(qc)\phi_N^{\nu+1}(c),\label{bl2}\\
%
%
%%%%%%%%%%%%%%% added for multiplicative formula
%
&& \phi_{N+1}^{\nu+1}(qc)\phi_{N}^{\nu}(c) - q^{2N}\phi_{N+1}^{\nu+1}(c)\phi_N^{\nu}(qc)
=c^2\left(a_0^2q^{2\nu}-q^{2N}\right)\phi_{N+1}^{\nu}(c)\phi_N^{\nu+1}(qc),\label{bl2:aux}\\
%%%%%%%%%%%%%%%%%%%%%%%%%%%%%%%%%%%%%%%%%%%%%
% typo corrected in bl3: sign of RHS
%%%%%%%%%%%%%%%%%%%%%%%%%%%%%%%%%%%%%%%%%%%%
&& \phi_N^{\nu+1}(qc)\phi_{N}^{\nu}(c) -\phi_{N}^{\nu+1}(c)\phi_N^{\nu}(qc)
=-\phi_{N-1}^{\nu}(qc)\phi_{N+1}^{\nu+1}(c),\label{bl3}\\
%
%%%%%%%%%%%%%%% added for multiplicative formula
%
&& \phi_N^{\nu+1}(qc)\phi_N^\nu(c)-q^{2N}\phi_N^{\nu+1}(c)\phi_N^\nu(qc)
=\frac{1}{a_0^2q^{2\nu}c^2}\phi_{N+1}^{\nu+1}(qc)\phi_{N-1}^{\nu}(c),\label{bl3:aux}\\
&&q^{2N}\phi_N^{\nu}(c)\phi_{N+1}^\nu(c) - \phi_{N+1}^{\nu+1}(c)\phi_N^{\nu-1}(c)
=-a_0^2c^2q^{2(\nu-1)}\phi_{N+1}^{\nu}(q^{-1}c)\phi_N^\nu(qc),\label{bl4}\\
&&\phi_N^{\nu}(c)\phi_{N+1}^{\nu+1}(c) +
c^2(a_0^2q^{2\nu}-q^{2N})\phi_{N+1}^{\nu}(c)\phi_N^{\nu+1}(c)
=q^{-2N}\phi_{N}^{\nu}(q^{-1}c)\phi_{N+1}^{\nu+1}(qc),\label{bl5}\\
&&a_0^2q^{2\nu}\phi_N^{\nu}(c)\phi_N^{\nu+1}(c)+\phi_{N+1}^{\nu+1}\phi_{N-1}^\nu(c) =
a_0^2q^{2\nu}\phi_N^{\nu+1}(q^{-1}c)\phi_N^{\nu}(qc).\label{bl6}
\end{eqnarray}
\end{prop}
{\it Proof of Theorem \ref{Toeplitz}.} We show that
$f_i(c)$ ($i=0,1,2$) defined by eqs.(\ref{f0})-(\ref{f2}) satisfy
eq.(\ref{qP4:1}). Substituting eqs.(\ref{f0})-(\ref{f2}) into the
numerator of right hand side of eq.(\ref{qP4:1}), we have
\begin{eqnarray*}
&& 1+q^{-N}f_2(c)+a_0q^{\nu-N}f_2(c)f_0(c) 
=1+q^{-N}f_2(c)\left(1+ a_0^2q^{2\nu}
c^2~\frac{\phi_{N+1}^\nu(c)\phi_N^\nu(qc)}{\phi_{N+1}^\nu(qc)\phi_N^\nu(c)}\right)\\
&=&1+q^{-N}f_2(c)~\frac{\phi_{N+1}^\nu(qc)\phi_N^\nu(c)+ a_0^2q^{2\nu}
c^2\phi_{N+1}^\nu(c)\phi_N^\nu(qc)}{\phi_{N+1}^\nu(qc)\phi_N^\nu(c)}\\
&=&1+q^{-N}f_2(c)~q^{-2N}\frac{\phi_{N+1}^{\nu+1}(qc)\phi_N^{\nu-1}(c)}{\phi_{N+1}^\nu(qc)\phi_N^\nu(c)}
=1-q^{-2N}\frac{\phi_{N}^{\nu-1}(qc)\phi_{N+1}^{\nu+1}(qc)}{\phi_{N}^\nu(qc)\phi_{N+1}^\nu(qc)}\\
&=& \frac{\phi_{N}^\nu(qc)\phi_{N+1}^\nu(qc)-q^{-2N}\phi_{N}^{\nu-1}(qc)\phi_{N+1}^{\nu+1}(qc)}
{\phi_{N}^\nu(qc)\phi_{N+1}^\nu(qc)}
=-a_0^2c^2q^{2(\nu-N)}\frac{\phi_{N+1}^{\nu}(c)\phi_N^\nu(q^2c)}{\phi_{N}^\nu(qc)\phi_{N+1}^\nu(qc)},
\end{eqnarray*}
where we used bilinear equations (\ref{bl1}) and (\ref{bl4}). 
The denominator is calculated by using eq.(\ref{bl2}) and (\ref{bl5}) as 
\begin{equation}
 1+a_0q^\nu f_0(c)+q^{N+1}f_0(c)f_1(c)=\frac{q^{-2N}\phi_{N}^{\nu-1}(c)\phi_{N+1}^{\nu}(q^2c)}{\phi_{N+1}^\nu(qc)\phi_N^{\nu-1}(qc)}.
\end{equation}
Then, the right hand side of eq.(\ref{qP4:1}) reduces to
\begin{eqnarray*}
&& q^{N+1}f_1(c)
\dfrac{1+q^{-N}f_2(c)+a_0q^{\nu-N}f_2(c)f_0(c)}{
1+a_0q^\nu f_0(c)+q^{N+1}f_0(c)f_1(c)} \\
&=& q^{N+1}\cdot \frac{1}{a_0q^{\nu+N-1}}~
\frac{\phi_{N+1}^\nu(qc)\phi_N^{\nu-1}(c)}{\phi_{N+1}^\nu(c)\phi_N^{\nu-1}(qc)}
\cdot a_0^2c^2q^{2(\nu-N)}\frac{\phi_{N+1}^{\nu}(c)\phi_N^\nu(q^2c)}{\phi_{N}^\nu(qc)\phi_{N+1}^\nu(qc)}
\cdot q^{2N}\frac{\phi_{N+1}^\nu(qc)\phi_N^{\nu-1}(qc)}{\phi_{N}^{\nu-1}(c)\phi_{N+1}^{\nu}(q^2c)}\\
&=& a_0c^2q^{\nu+2}~\frac{\phi_{N+1}^\nu(qc)\phi_N^\nu(q^2c)}{\phi_{N+1}^{\nu}(q^2c)\phi_{N}^\nu(qc)}=f_0(qc).
\end{eqnarray*}
Thus we have shown that eq.(\ref{qP4:1}) follows. Eq.(\ref{qP4:2}) is checked by using the
bilinear equations (\ref{bl2}), (\ref{bl4}), (\ref{bl3})
and(\ref{bl6}). Eq.(\ref{qP4:3}) follows automatically.\hfill\qed
\par\medskip
\noindent\textbf{Remark.} It should be noted that both $1+a_if_i$ and $1+f_i/a_i$ ($i=0,1,2$)
 admit multiplicative formula with respect to $\phi$. In fact, we have,
%%%%%%%%%%%%%%%%%%%%%%%%%%%%%%%%%%%%%%%%%%
% typo in 3rd eq. corrected: sign of RHS
%%%%%%%%%%%%%%%%%%%%%%%%%%%%%%%%%%%%%%%%%%
\begin{eqnarray}
&&1+a_0q^\nu f_0(c;\nu,N)=q^{-2N}\frac{\phi_{N+1}^{\nu+1}(qc)\phi_N^{\nu-1}(c)}
{\phi_{N+1}^{\nu}(qc)\phi_{N}^\nu(c)},\\
&&1+a_1q^{-\nu+N}f_1(c;\nu,N)=\frac{a_0^2q^{2(\nu-1)}-q^{2N}}{a_0^2q^{2(\nu-1)}}
\frac{\phi_{N+1}^{\nu-1}(qc)\phi_N^{\nu}(c)}{\phi_{N+1}^{\nu}(c)\phi_{N}^{\nu-1}(qc)},\\
&&1+a_2q^{-N}f_2(c;\nu,N)=
-\frac{\phi_{N-1}^{\nu-1}(qc)\phi_{N+1}^{\nu}(c)}{\phi_N^{\nu-1}(c)\phi_N^\nu(qc)},
\end{eqnarray}
and
%%%%%%%%%%%%%%%%%%%%%%%%%%%%%%%%%%%%%%%%%%%%%%%%%%%%%%%%%%%%%%%
% typo in 2nd and 3rd eq. corrected: assignment of c and qc
%%%%%%%%%%%%%%%%%%%%%%%%%%%%%%%%%%%%%%%%%%%%%%%%%%%%%%%%%%%%%%%%
\begin{eqnarray}
&& 1+\frac{f_0(c;\nu,N)}{a_0q^\nu}=
\frac{\phi_{N}^{\nu-1}(qc)\phi_{N+1}^{\nu+1}(c)}{\phi_{N+1}^{\nu}(qc)\phi_N^\nu(c)},\\
&& 1+\frac{f_1(c;\nu,N)}{a_1q^{-\nu+N}}=q^{-2N}c^2(q^{2N}-a_0^2q^{2(\nu-1)})
\frac{\phi_{N+1}^{\nu-1}(c)\phi_N^\nu(qc)}
{\phi_{N+1}^\nu(c)\phi_N^{\nu-1}(qc)},\\
&&1+\frac{f_2(c;\nu,N)}{a_2q^{-N}} =\frac{1}{a_0^2c^2q^{2(\nu-1)}}
\frac{\phi_{N+1}^\nu(qc)\phi_{N-1}^{\nu-1}(c)}
{\phi_N^\nu(qc)\phi_N^{\nu-1}(c)},
\end{eqnarray}
which can be verified directly by using eqs.(\ref{f0})-(\ref{f2}) and the bilinear equations
(\ref{bl1})-(\ref{bl3:aux}). 
%These multiplicative formulas provide us
%with an important key for introducing $\tau$ functions in general setting,
%which will be discussed in Section 3.

\subsubsection{Proof of Proposition \ref{bilinear}}
Our basic idea for proving Proposition \ref{bilinear} is as
follows. Bilinear difference equations are derived from the Pl\"ucker
relations, which are quadratic identities among determinants whose
columns are shifted. Therefore, we first construct such  ``difference
formulas'' that relate ``shifted determinants'' and $\phi_N^\nu(c)$, by
using the contiguity relations of $G_\nu(c)$.  
We then derive bilinear difference equations
with the aid of difference formulas from proper Pl\"ucker relations.
We take eq.(\ref{bl1}) as an example to show this procedure explicitly.
For other equations, see Appendix A. 

Let us introduce notations,
\begin{equation}
 G_\nu(q^mc)=G_\nu^m,\quad  \phi_N^\nu(q^mc)=\phi_N^{\nu,m},\label{Gnm}
\end{equation}
\begin{equation}
\phi_N^{\nu,m}=\left|
\begin{array}{cccc}
 G_\nu^m&G_{\nu+1}^m &\cdots &G_{\nu+N-1}^m \\
 G_{\nu-1}^m&G_{\nu}^m &\cdots &G_{\nu-N-2}^m \\
\vdots & \vdots & \ddots & \vdots\\
 G_{\nu-N+1}^m&G_{\nu-N+2}^m &\cdots &G_{\nu}^m
\end{array}\right| 
=\left|\mathbf{0}^m,\ \mathbf{1}^m,\ \cdots,\ \mathbf{N-1}^m\right|,\label{phi:1}
\end{equation}
where $\mathbf{k}^m$ denotes a column vector,
\begin{equation}
 \mathbf{k}^m=\left(\begin{array}{c}G_{\nu+k}^m \\G_{\nu+k-1}^m\\\vdots\\G_{\nu+k-N+1}^m
		  \end{array}\right).
\end{equation}
Here the height of the column is $N$, but we use the same symbol for
determinants with different size, since there is no possibility of
confusion. 

We next construct a difference formula.
\begin{lem}\label{Difference Formula I:lem}{\bf (Difference Formula I)}
\begin{equation}
 |\mathbf{0}^{m+1\prime},\ \mathbf{0}^m,\ \mathbf{1}^m,\cdots,\ \mathbf{N-3}^m,\
 \mathbf{N-2}^m| = a_0^{-2(N-1)}c^{-2(N-1)}q^{-2(N-1)(m+\nu-1)}~\phi_N^{\nu,m+1},\label{diff Formula I(1)}
\end{equation}
 \begin{equation}
 |\mathbf{1}^{m+1\prime},\ \mathbf{0}^m,\ \mathbf{1}^m,\cdots,\ \mathbf{N-3}^m,\
 \mathbf{N-2}^m| = a_0^{-2(N-1)}c^{-2(N-1)}q^{-2(N-1)(m+\nu-1)}~\phi_N^{\nu,m+1},\label{diff Formula I(2)}
\end{equation} 
where $\mathbf{k}^{m\prime}$ is a column vector,
\begin{equation}
\mathbf{k}^{m\prime}=\left(\begin{array}{c}G_{\nu+k}^m \\q^2G_{\nu+k-1}^m\\\vdots\\q^{2(N-1)}G_{\nu+k-N+1}^m
		  \end{array}\right). 
\end{equation}
\end{lem}
\noindent{\it Proof of Lemma \ref{Difference Formula I:lem}.} We use the contiguity
relation (\ref{contiguity:cn1}), which is rewritten as
\begin{equation}
 G_{\nu+1}^{m+1}=G_{\nu}^{m+1}+a_0^2q^{2(\nu+m)}c^2G_\nu^m.\label{contiguity:dn1}
\end{equation}
in the present notation.
Then we have
\begin{eqnarray*}
\phi_N^{\nu,m+1}&=& \left|\begin{array}{cccc}
 G_\nu^{m+1}&G_{\nu+1}^{m+1} &\cdots &G_{\nu+N-1}^{m+1} \\
 G_{\nu-1}^{m+1}&G_{\nu}^{m+1} &\cdots &G_{\nu-N-2}^{m+1} \\
\vdots & \vdots & \ddots & \vdots\\
 G_{\nu-N+1}^{m+1}&G_{\nu-N+2}^{m+1} &\cdots &G_{\nu}^{m+1}
\end{array}\right| \\ 
&=& \left|\begin{array}{cccc}
 G_\nu^{m+1}    &G_{\nu+1}^{m+1}-G_\nu^{m+1}  &\cdots &G_{\nu+N-1}^{m+1}-G_{\nu+N-2}^{m+1} \\
 G_{\nu-1}^{m+1}&G_{\nu}^{m+1}-G_{\nu-1}^{m+1}&\cdots &G_{\nu+N-2}^{m+1}-G_{\nu+N-3}^{m+1} \\
 \vdots         & \vdots                      & \ddots& \vdots\\
 G_{\nu-N+1}^{m+1}&G_{\nu-N+2}^{m+1}- G_{\nu-N+1}^{m+1} &\cdots &G_{\nu}^{m+1} - G_{\nu-1}^{m+1}
\end{array}\right| \\ 
&=&\left|\begin{array}{cccc}
 G_\nu^{m+1}    &a_0^2c^2q^{2(\nu+m)}G_\nu^{m}      &\cdots &a_0^2q^{2(\nu+N-2+m)}c^2G_{\nu+N-2}^{m} \\
 G_{\nu-1}^{m+1}&a_0^2c^2q^{2(\nu-1+m)}G_{\nu-1}^{m}&\cdots &a_0^2q^{2(\nu+N-3+m)}c^2G_{\nu+N-3}^{m} \\
 \vdots         & \vdots                      & \ddots& \vdots\\
 G_{\nu-N+1}^{m+1}&a_0^2c^2q^{2(\nu-N+1+m)}G_{\nu-N+1}^{m+1} &\cdots &a_0^2q^{2(\nu-1+m)}c^2G_{\nu-1}^{m}
\end{array}\right| \\ 
&=&a_0^{2(N-1)}c^{2(N-1)}q^{2(N-1)(\nu+m-1)}\left|\begin{array}{cccc}
 G_\nu^{m+1}    &G_\nu^{m}      &\cdots &G_{\nu+N-2}^{m} \\
 q^2G_{\nu-1}^{m+1}&G_{\nu-1}^{m}&\cdots &G_{\nu+N-3}^{m} \\
 \vdots         & \vdots                      & \ddots& \vdots\\
 q^{2(N-1)}G_{\nu-N+1}^{m}&G_{\nu-N+1}^{m} &\cdots &G_{\nu-1}^{m}
\end{array}\right|  \\
&=&a_0^{2(N-1)}c^{2(N-1)}q^{2(N-1)(\nu+m-1)}~|\mathbf{0}^{m+1\prime},\ \mathbf{0}^m,\ \mathbf{1}^m,\cdots,\ \mathbf{N-3}^m,\
 \mathbf{N-2}^m|,
\end{eqnarray*}
which is eq.(\ref{diff Formula I(1)}). In the second line of the above
calculation, adding the second column to the first column, we get
\begin{equation}
\phi_N^{\nu,m+1}=\left|\begin{array}{cccc}
 G_{\nu+1}^{m+1}    &G_{\nu+1}^{m+1}-G_\nu^{m+1}  &\cdots &G_{\nu+N-1}^{m+1}-G_{\nu+N-2}^{m+1} \\
 G_{\nu}^{m+1}&G_{\nu}^{m+1}-G_{\nu-1}^{m+1}&\cdots &G_{\nu-N-2}^{m+1}-G_{\nu+N-3}^{m+1} \\
 \vdots         & \vdots                      & \ddots& \vdots\\
 G_{\nu-N+2}^{m+1}&G_{\nu-N+2}^{m+1}- G_{\nu-N+1}^{m+1} &\cdots &G_{\nu}^{m+1} - G_{\nu-1}^{m+1}
\end{array}\right|  ,
\end{equation}
{}from which we obtain eq.(\ref{diff Formula I(2)}) by using
eq.(\ref{contiguity:dn1}).\hfill\qed\medskip

We then consider the Pl\"ucker relation,
\begin{eqnarray}
  0&=&\left|\varphi_1,\ \mathbf{1}^{m+1\prime},\ \mathbf{1}^{m},\cdots,\mathbf{N-1}^{m}\right|
\times\left|\mathbf{0}^{m},\ \mathbf{1}^{m},\cdots,\mathbf{N-1}^{m},\ \mathbf{N}^{m}\right|\nonumber\\
   &-&\left|\varphi_1,\ \mathbf{0}^{m},\ \mathbf{1}^{m},\cdots,\mathbf{N-1}^{m}\right|
\times\left|\mathbf{1}^{m+1\prime},\ \mathbf{1}^{m},\cdots,\mathbf{N-1}^{m},\ \mathbf{N}^{m}\right|\nonumber\\
   &+&\left|\mathbf{1}^{m+1\prime},\ \mathbf{0}^{m},\ \mathbf{1}^{m},\cdots,\mathbf{N-1}^{m}\right|
\times\left|\varphi_1,\ \mathbf{1}^{m},\cdots,\mathbf{N-1}^{m},\ \mathbf{N}^{m}\right|,\label{pl10}
\end{eqnarray}
where
 \begin{equation}
 \varphi_1=\left(\begin{array}{c}1\\0\\ \vdots\\0 \end{array}\right).
\end{equation}
By expansion with respect to the column $\varphi_1$, the Pl\"ucker
relation (\ref{pl10}) is rewritten as
\begin{eqnarray}
0&=&q^2\left|\mathbf{0}^{m+1\prime},\ \mathbf{0}^{m},\cdots,\mathbf{N-2}^{m},\ \mathbf{N-1}^{m}\right|
\times\left|\mathbf{0}^{m},\ \mathbf{1}^{m},\cdots,\mathbf{N-1}^{m}\right|\nonumber\\
   &-&\left|\mathbf{-1}^{m},\ \mathbf{0}^{m},\cdots,\mathbf{N-2}^{m},\ \mathbf{N-1}^{m}\right|
\times\left|\mathbf{1}^{m+1\prime},\ \mathbf{1}^{m},\cdots,\mathbf{N-1}^{m}\right|\nonumber\\
   &+&\left|\mathbf{1}^{m+1\prime},\ \mathbf{0}^{m},\ \mathbf{1}^{m},\cdots,\mathbf{N-1}^{m}\right|
\times\left|\mathbf{0}^{m},\cdots,\mathbf{N-2}^{m},\ \mathbf{N-1}^{m}\right|,\nonumber
\end{eqnarray}
where we have used
\begin{eqnarray*}
&& \left|\varphi_1,\ \mathbf{1}^{m+1\prime},\ \mathbf{1}^{m},\cdots,\mathbf{N-1}^{m}\right|
=\left|
\begin{array}{ccccc}
 1& G_{\nu+1}^{m+1}&G_{\nu+1}^m &\cdots&G_{\nu+N-1}^m \\
 0& q^2G_{\nu}^{m+1}&G_{\nu}^m &\cdots&G_{\nu+N-2}^m \\
 \vdots& \vdots & \vdots & \ddots & \vdots\\
 0& q^{2N}G_{\nu-N+1}^{m+1}&G_{\nu-N+2}^m &\cdots&G_{\nu-1}^m 
\end{array}
\right|\\
&=&\left|
\begin{array}{cccc}
q^2G_{\nu}^{m+1}&G_{\nu}^m &\cdots&G_{\nu+N-2}^m \\
\vdots & \vdots & \ddots & \vdots\\
q^{2N}G_{\nu-N+1}^{m+1}&G_{\nu-N+2}^m &\cdots&G_{\nu-1}^m 
\end{array}
\right|
=q^2\left|\mathbf{0}^{m+1\prime},\ \mathbf{0}^{m},\cdots,\mathbf{N-2}^{m},\ \mathbf{N-1}^{m}\right|.
\end{eqnarray*}
Then, we obtain by using Lemma \ref{Difference Formula I:lem},
\begin{eqnarray}
0&=&q^2\times a_0^{-2(N-1)}c^{-2(N-1)}q^{-2(N-1)(\nu+m-1)}~\phi_N^{\nu,m+1}\times \phi_{N+1}^{\nu,m}\nonumber\\
&-&\phi_N^{\nu-1,m}\times a_0^{-2N}c^{-2N}q^{-2N(\nu+m)}~\phi_{N+1}^{\nu+1,m+1}\nonumber\\
&+& a_0^{-2N}c^{-2N}q^{-2N(\nu+m-1)}~\phi_{N+1}^{\nu,m+1}\times \phi_N^{\nu,m},
\end{eqnarray}
which is the same as eq.(\ref{bl1}). This completes the proof of the
bilinear equation (\ref{bl1}).

\subsubsection{Determinant formula for negative $N$}
Proposition \ref{Toeplitz} is a determinant formula for the solutions with
$N\in \BZ_{\ge 0}$, which are obtained by successive application of $T_2$ on the seed
solution described in Proposition \ref{seed}. In order to obtain
solutions for $N\in\BZ_{< 0}$, we have to apply $T_2^{-1}$ on the seed
solution, however, we find that this procedure collapses. Therefore, we
construct ``seed and higher solutions'' for negative $N$ so that bilinear
equations described in Proposition \ref{bilinear} hold for all 
$N \in \BZ_{<0}$. 
To find the seed solutions for $N=-1$, we put $N=-1$,  
$\phi_{-1}^\nu(c)=\overline{G}_\nu(c)$ and $\phi_0^\nu(c)=1$ in
eqs.(\ref{bl1})-(\ref{bl6}), we obtain a set of contiguity relations to
be satisfied by $\overline{G}_\nu(c)$,
\begin{eqnarray}
&& a_0^2c^2q^{2\nu}\overline{G}_{\nu}(qc)+ \overline{G}_\nu(c)
=q^{2}\overline{G}_{\nu-1}(c),\label{qHermitebar:cn1}\\
&& \overline{G}_{\nu}(qc)=\overline{G}_{\nu+1}(c)+c^2\overline{G}_{\nu+1}(qc).\label{qHermitebar:cn2}
\end{eqnarray}
Similarly to the case of $N\geq 0$, the following 
three-term relations in the direction of $c$ or $\nu$ are derived from
eqs.(\ref{qHermitebar:cn1}) and (\ref{qHermitebar:cn2}),
%\begin{eqnarray}
%&& a_0^2q^{2\nu}\overline{G}_{\nu}(qc) -\overline{G}_{\nu}(c)
%=(a_0^2q^{2\nu}-q^{-2})\overline{G}_{\nu+1}(c),\\
%%
%&& \overline{G}_{\nu}(c) - q^{-2}\overline{G}_{\nu}(qc)
%=c^2\left(a_0^2q^{2\nu}-q^{-2}\right)\overline{G}_{\nu+1}(qc),
%\end{eqnarray}
%or
\begin{eqnarray}
 &&\overline{G}_{\nu}(c) = \left(\frac{1}{q^2}-c^2\right)\overline{G}_{\nu}(qc)
+a_0^2q^{2\nu}c^2\overline{G}_{\nu}(q^2c), \label{qHermitebar:c}\\
 &&\overline{G}_{\nu}(c) = \frac{1}{q^2}\left(1+c^2\right)\overline{G}_{\nu+1}(c)
-\frac{c^2}{q^4}\left(1-a_0^2q^{2\nu}\right)\overline{G}_{\nu+2}(c).
\label{qHermitebar:nu}
\end{eqnarray}
Then we obtain a determinant formula for $N<0$ as follows.
\begin{thm}\label{Toeplitz:N<0}
 Let $\overline{G}_\nu(c)$ is a function satisfying 
eqs. (\ref{qHermitebar:cn1}) and (\ref{qHermitebar:cn2}).  
For each $-M=N\in\BZ_{< 0}$, 
we define an $M\times M$
Toeplitz determinant 
\begin{equation}
 \phi_N^\nu(c)=
\det \left(\overline{G}_{\nu-i+j}(c)\right)_{i,j=1,\cdots,M}
\quad (-M=N\in\BZ_{<0}),
\end{equation}
Then, $\phi_N^\nu(c)$ satisfies the bilinear equations
 (\ref{bl1})-(\ref{bl6}), and therefore
eqs.(\ref{f0})-(\ref{f2}) satisfy
$q$P$_{\IV}$(\ref{qP4:1})-(\ref{qP4:3}) with $a_2=1$.
\end{thm}
Since this theorem is proved by similar procedure to Theorem
\ref{Toeplitz}, we omit the details.
\par\medskip
\noindent\textbf{Remark}
If we parametrize as $a_0=q^\mu$, 
\begin{equation}
 \overline{G}_\nu(c;q)=c^2q^{-2\nu} G_{-(\nu-2\mu)}(c;1/q),
\end{equation}
satisfies eqs.(\ref{qHermitebar:cn1}) and (\ref{qHermitebar:cn2}) formally.

\subsection{$q$-Okamoto polynomials}
In the previous sections we have discussed the particular solutions
on the reflection hyperplane of the affine Weyl group $W=W(A^{(1)}_2)$
in the parameter space. It is easy to see that $q$P$_{\IV}$ (\ref{qP4})
with normalization condition (\ref{normalization}) admits a particular
solution given by
\begin{equation}
(f_0,f_1,f_2;a_0,a_1,a_2)=(x p^{-1},x p^{-1},x p^{-1};p^{-1},p^{-1},p^{-1}),\quad x=c^{2/3},\quad p=q^{-1/3}.
\end{equation}
Note that this parameter corresponds to the fixed point of $\pi$, namely
Dynkin diagram automorphism of the affine Weyl group $W(A^{(1)}_2)$.
By applying the B\"acklund transformations on this seed solution, we get
a series of rational solutions on the barycenters of the Weyl chamber of
the affine Weyl group $W(A^{(1)}_2)$. 
\begin{thm}\label{qOkamoto}
For the parameters
\begin{equation}
T_1^m T_2^n(a_0,a_1,a_2)=(p^{-3m-1}, p^{3m-3n-1},p^{3n-1}),
\quad m,n \in \BZ,
\end{equation}
we have the following rational solutions of {\rm $q$P$_{\IV}$}
equation (\ref{qP4}),
\begin{equation}
\begin{array}l
T_1^m T_2^n(f_0)=
x p^{m-2 n-1}\displaystyle\frac{Q_{m+1,n}(x p^{-2})Q_{m+1,n+1}(x)}
{Q_{m+1,n}(x)Q_{m+1,n+1}(x p^{-2})}, \\[4mm]
T_1^mT_2^m(f_1)=
x p^{m+n-1}\displaystyle\frac{Q_{m+1,n+1}(x p^{-2})Q_{m,n}(x)}
{Q_{m+1,n+1}(x)Q_{m,n}(x p^{-2})}, \\[4mm]
T_1^mT_2^n(f_2)=
x p^{n-2 m-1}\displaystyle\frac{Q_{m,n}(x p^{-2})Q_{m+1,n}(x)}
{Q_{m,n}(x)Q_{m+1,n}(x p^{-2})}.
\end{array}
\end{equation}
Here $Q_{m,n}(x)$ are monic polynomials defined through the recurrence
relations,
\begin{eqnarray*}\label{Okamoto:bl}
Q_{m-1,n}(x)Q_{m+1,n+1}(x p^{-2})&=&
Q_{m,n}(x)Q_{m,n+1}(x p^{-2})+
x p^{-2m-2n}Q_{m,n+1}(x)Q_{m,n}(x p^{-2}), \\[1mm]
Q_{m+1,n}(x)Q_{m,n+1}(x p^{-2})&=&
Q_{m+1,n+1}(x)Q_{m,n}(x p^{-2})+
x p^{4m-2n-2}Q_{m,n}(x)Q_{m+1,n+1}(x p^{-2}), \\[1mm]
Q_{m+1,n+1}(x)Q_{m,n-1}(x p^{-2})&=&
Q_{m,n}(x)Q_{m+1,n}(x p^{-2})+
x p^{4n-2m-2}Q_{m+1,n}(x)Q_{m,n}(x p^{-2}),
\end{eqnarray*}
with initial conditions
$Q_{0,0}(x)=Q_{1,0}(x)=Q_{1,1}(x)=1$.
\end{thm}
We call the polynomials $Q_{m,n}(x)$ the {\it $q$-Okamoto polynomials}. 
Some examples are as follows.
\begin{eqnarray*}
Q_{2,0}(x) &=& x^2+(p^2+1)x+1,\\
Q_{3,0}(x) &=& x^6 + x^5 (p^6 + p^4 + p^2 + 2 + p^{-2} ) + 
x^4 (2 p^8 + 2 p^6 + 3 p^4 + 3 p^2 + 3 + p^{-2} + p^{-4} )\\
&& + x^3 (p^{10} + 2 p^8 + 3 p^6 + 4 p^4 + 4 p^2  + 3 + 2 p^{-2} + p^{-4})\\
&&+ x^2 (2 p^8 + 2 p^6 + 3 p^4 + 3 p^2 + 3 + p^{-2} + p^{-4})
 + x (p^6 + p^4 + p^2 + 2 + p^{-2} ) + 1,\\
Q_{2,1}(x) &=& x+1,\\
Q_{3,1}(x) &=& x^4 + x^3(p^4 + p^2 + 1 + p^{-2} ) + 
x^2(2p^4 + p^2 + 2 + p^{-2} )+ x(p^4 + p^2 + 1 + p^{-2} ) + 1.
\end{eqnarray*}
Similarly to the continuous case,
$q$-Okamoto polynomials admit determinant formula of Jacobi-Trudi type.
The proof of the Theorem \ref{qOkamoto}
as well as the Jacobi-Trudi type formulas
will be given in our next paper\cite{KNY2}, where we will discuss
the $\tau$ functions in a more general setting.
%%%%%%%%%%%%%%%%%%%%%%%%%%%%%%%%%
\def\comentout#1{{}}
\comentout{
In the previous sections we have discussed the particular solutions
on the reflection hyperplane of the affine Weyl group $W=W(A^{(1)}_2)$
in the parameter space. It is easy to see that $q$P$_{\IV}$ (\ref{qP4})
with normalization condition (\ref{normalization}) admit a particular
solution given by
\begin{equation}
 (f_0,f_1,f_2;a_0,a_1,a_2)=(x/p,x/p,x/p;1/p,1/p,1/p),\quad x=c^{2/3},\quad p=q^{-1/3}.
\end{equation}
Note that this parameter corresponds to the fixed point of $\pi$, namely
Dynkin diagram automorphism of the affine Weyl group $W(A^{(1)}_2)$.
By applying the B\"acklund transformations on this seed solution, we get
a series of rational solutions on the barycenters of the Weyl chamber of
the affine Weyl group $W(A^{(1)}_2)$. Moreover, we observe that those
rational solutions are factorized in terms of two sequences of polynomials with
shifted $x$.  For example, applying $T_1$ and $T_1^2$ on the above seed
solutions, we obtain,
\begin{equation}
 \begin{array}{lll}
\smallskip
  T_1(f_0)=x~\dfrac{Q_2(x/p^2)R_1(x)}{Q_2(x)R_1(x/p^2)}, & 
  T_1(f_1)=x~\dfrac{Q_1(x)R_1(x/p^2)}{Q_1(x/p^2)R_1(x)},& 
  T_1(f_2)=p^{-3}x~\dfrac{Q_1(x/p^2)Q_2(x)}{Q_1(x)Q_2(x/p^2)},\\
  T_1^2(f_0)=px~\dfrac{Q_3(x/p^2)R_2(x)}{Q_3(x)R_2(x/p^2)}, & 
  T_1^2(f_1)=px~\dfrac{Q_2(x)R_2(x/p^2)}{Q_2(x/p^2)R_2(x)},& 
  T_1^2(f_2)=p^{-5}x~\dfrac{Q_2(x/p^2)Q_3(x)}{Q_2(x)Q_3(x/p^2)},\\
 \end{array}
\end{equation}
respectively, where 
\begin{eqnarray*}
Q_1(x) &=& 1,\\
Q_2(x) &=& x^2+(p^2+1)x+1,\\
Q_3(x) &=& x^6 + x^5 (p^6 + p^4 + p^2 + 2 + p^{-2} ) + x^4 (2 p^8 + 2 p^6 + 3 p^4 
+ 3 p^2 + 3 + p^{-2} + p^{-4} )\\
&& + x^3 (p^{10} + 2 p^8 + 3 p^6 + 4 p^4 + 4 p^2  + 3 + 2 p^{-2} + p^{-4})\\
&&+ x^2 (2 p^8 + 2 p^6 + 3 p^4 + 3 p^2 + 3 + p^{-2} + p^{-4})
 + x (p^6 + p^4 + p^2 + 2 + p^{-2} ) + 1,\\
% R_0(x)&=&1,\\
 R_1(x) &=& x+1,\\
 R_2(x) &=& x^4 + x^3(p^4 + p^2 + 1 + p^{-2} ) + x^2(2p^4 + p^2 + 2 + p^{-2} )
+ x(p^4 + p^2 + 1 + p^{-2} ) + 1.
\end{eqnarray*}
We call such polynomials that characterize this class of rational
solutions {\it $q$-Okamoto polynomials}. Similarly to the continuous case,
$q$-Okamoto polynomials admit determinant formula of Jacobi-Trudi type.
This formula is derived by considering the $\tau$ function in the
general setting, which will be discussed in the next paper.\cite{KNY2}
}

\section{Discussion}
\subsection{$\widetilde{W}(A^{(1)}_2) \times \widetilde{W}(A^{(1)}_1)$ symmetry
and comparison with Sakai's Mul.6 system}

Along with the transformations $s_i$ and $\pi$ of $\widetilde W(A^{(1)}_2)$,
we define the transformations $w_0$, $w_1$ and $r$ acting on
$\BC(a_i,f_i\ (i=0,1,2))$ as follows:
\begin{equation}\label{A11action}
\begin{array}l
w_0(f_0)=\dfrac{a_0a_1(a_2 a_0+a_2 f_0+f_2 f_0)}{f_2(a_0 a_1+a_0 f_1+f_0 f_1)},\quad
w_1(f_0)=\dfrac{1+a_0 f_0+a_0 a_1 f_0 f_1}{a_0 a_1 f_1(1+a_2 f_2+a_2 a_0 f_2 f_0)},\\[3mm]
w_0(f_1)=\dfrac{a_1a_2(a_0 a_1+a_0 f_1+f_0 f_1)}{f_0(a_1 a_2+a_1 f_2+f_1 f_2)},\quad
w_1(f_1)=\dfrac{1+a_1 f_1+a_1 a_2 f_1 f_2}{a_1 a_2 f_2(1+a_0 f_0+a_0 a_1 f_0 f_1)},\\[3mm]
w_0(f_2)=\dfrac{a_2a_0(a_1 a_2+a_1 f_2+f_1 f_2)}{f_1(a_2 a_0+a_2 f_0+f_2 f_0)},\quad
w_1(f_2)=\dfrac{1+a_2 f_2+a_2 a_0 f_2 f_0}{a_2 a_0 f_0(1+a_1 f_1+a_1 a_2 f_1 f_2)},\\[3mm]
r(f_i)=\dfrac{1}{f_i}, \quad 
w_0(a_i)=w_1(a_i)=r(a_i)=a_i, \quad
(i=0,1,2). 
\end{array}
\end{equation}

\begin{lem}
The transformations $w_0$, $w_1$ and $r$ generate the extended affine Weyl group 
$\widetilde{W}(A^{(1)}_1)$,
namely we have
\begin{equation}
w_0^2=w_1^2=r^2=1, \quad r w_0=w_1 r.
\end{equation}
Moreover, this action of $\widetilde{W}(A^{(1)}_1)=\langle w_0, w_1, r \rangle$ 
commutes with that of $\widetilde{W}(A^{(1)}_2)=\langle s_0, s_1, s_2, \pi \rangle$.
\end{lem}

Note that the discrete time evolution of the $q P_{IV}$ system is a translation
of the $\widetilde{W}(A^{(1)}_1)$, that is, 
\begin{equation}
t=T_4=r w_0.
\end{equation}

\begin{lem}
The representation of $\widetilde{W}(A^{(1)}_2) \times \widetilde{W}(A^{(1)}_1)$
is equivalent to Sakai's system Mul.6.
\end{lem}
\def\ta{\tilde{a}}
\def\tf{\tilde{f}}
{\it Proof.} 
Let $x, y, z$ and $\ta_0, \ta_1, \ta_2, b_1, b_0=\ta_0 \ta_1 \ta_2/b_1$ be 
Sakai's homogeneous variables and the parameters\cite{Sak}.
Then his representation of $\widetilde{W}(A^{(1)}_2) \times \widetilde{W}(A^{(1)}_1)$
is given as follows:
\begin{equation}
\pi=\sigma^4, \quad
s_2=\pi s_1 \pi^2, \quad
s_0=\pi^2 s_1 \pi, \quad
r=\sigma^3, \quad
w_0=r w_1 r.
\end{equation}
Where $\sigma$, $s_1$ and $w_1$ is given by,
\footnote{The generators $\sigma, s_1, w_1$ here correspond with 
$\sigma_{(123450)}, w_1$ and $w'_1$ in \cite{Sak}.
An error in the formula of $\sigma_{(123450)}$ \cite{Sak} is corrected.}
\begin{equation}
\begin{array}l
\sigma(\ta_0)=\ta_1,\quad
\sigma(\ta_1)=\ta_2,\quad
\sigma(\ta_2)=\ta_0,\quad
\sigma(b_0)=b_1,\quad
\sigma(b_1)=b_0,\\[1mm]
\sigma(x)=\ta_2 x y(z-x),\quad
\sigma(y)=-b_1 y z(x+y-z),\quad
\sigma(z)=\ta_2 x(x-z)^2,\\[3mm]
s_1(\ta_0)=\ta_0 \ta_1, \quad
s_1(\ta_1)=\ta_1^{-1}, \quad
s_1(\ta_2)=\ta_2 \ta_1, \quad
s_1(b_i)=b_i,\quad (i=0,1)\\[1mm]
s_1(x)=x, \quad
s_1(y)=\ta_1 y, \quad
s_1(z)=\ta_1 z,\\[3mm]
w_1(\ta_i)=\ta_i, \quad (i=0,1,2) \quad
w_1(b_0)=b_0 b_1^2,\quad
w_1(b_1)=b_1^{-1},\\[1mm]
w_1(x)=x(y+\ta_2 x), \quad
w_1(y)=-b_1 y(y+\ta_2 x), \quad
w_1(z)=z(\ta_2 x-b_1 y).
\end{array}
\end{equation}
Introduce variables $\tf_i$ $(i=0,1,2)$ as
\begin{equation}
\tf_1=-\dfrac{x}{z}, \quad
\tf_2=\ta_2 \dfrac{x-z}{y}, \quad
\tf_0=\dfrac{b_1 y z}{\ta_2 x(x-z)}.
\end{equation}
Then we have
\begin{equation}
\begin{array}l
\sigma(\tf_0)=\dfrac{\ta_1}{\tf_1},\quad
\sigma(\tf_1)=\dfrac{\ta_2}{\tf_2},\quad
\sigma(\tf_2)=\dfrac{\ta_0}{\tf_0},\\[3mm]
s_1(\tf_0)=\tf_0 \dfrac{\ta_1(1+\tf_1)}{\ta_1+\tf_1}, \quad
s_1(\tf_1)=\dfrac{\tf_1}{\ta_1}, \quad
s_1(\tf_2)=\dfrac{\ta_1+\tf_2}{1+\tf_1},\\[3mm]
w_1(\tf_0)=\dfrac{1+\tf_0+\tf_0 \tf_1}{\tf_1(1+\tf_2+\tf_2 \tf_0)}\quad
w_1(\tf_1)=\dfrac{1+\tf_1+\tf_1 \tf_2}{\tf_2(1+\tf_0+\tf_0 \tf_1)}\quad
w_1(\tf_2)=\dfrac{1+\tf_2+\tf_2 \tf_0}{\tf_0(1+\tf_1+\tf_1 \tf_2)}.
\end{array}
\end{equation}
This representation is equivalent to our representation
by the relation $\tf_i=a_i f_i, \quad \ta_i=a_i^2$. \hfill\qed
\subsection{Ultra-discretization of $q$P$_{\IV}$} 
In Section 1, we have mentioned that $q$P$_{\IV}$ admit a continuous
limit to the symmetric form of P$_{\IV}$. There is another interesting
limit, which is known as ``ultra-discrete limit''\cite{ultra}. We put
\begin{displaymath}
 f_i={\rm e}^{F_i/\varepsilon},\quad a_i={\rm e}^{A_i/\varepsilon},\quad i=0,1,2
\end{displaymath}
and take the limit $\varepsilon\to +0$. By using the formula,
\begin{equation}
 \lim_{\varepsilon\to +0}\varepsilon\log\left({\rm e}^{\frac{A}{\varepsilon}}
+{\rm e}^{\frac{B}{\varepsilon}}+\cdots\right)
=\max(A,B,\cdots),
\end{equation}
$q$P$_{\IV}$(\ref{qP4}) yields
\begin{eqnarray}
 \overline{F}_0=t(F_0)=A_0+A_1+F_1&+&\max\left(0,\ A_2+F_2,\ A_0+A_2+F_0+F_2\right)\nonumber\\
&-&\max\left(0,\ A_0+F_0,\ A_1+A_0+F_1+F_0\right),\nonumber\\
 \overline{F}_1=t(F_1)=A_1+A_2+F_2&+&\max\left(0,\ A_0+F_0,\ A_1+A_0+F_1+F_0\right)\nonumber\\
&-&\max(0,\ A_1+F_1,\ A_2+A_1+F_2+F_1),\\
 \overline{F}_2=t(F_2)=A_2+A_0+F_0&+&\max\left(0,A_1+F_1,\ A_2+A_1+F_2+F_1\right)\nonumber\\
&-&\max\left(0,A_2+F_2,\ A_0+A_2+F_0+F_2\right),\nonumber\\
\overline{A}_i=t(A_i)=A_i, \nonumber
\end{eqnarray}
which we call the fourth ultra-discrete Painlev\'e equation(uP$_{\IV}$).
Simultaneously, B\"acklund transformation (\ref{W:general}) is ultra-discretized as
\begin{equation}
\begin{array}{l}\label{ultraW}
\smallskip
s_i(F_j)=F_j+u_{ij}\left(\max(A_i,\ F_i)-\max(0,A_i+F_i)\right),\\
s_i(A_j)=A_j-a_{ij}A_i,\quad \pi(X_i)=X_{i+1}\quad (i=0,1,2) ,\quad X=F,A 
\end{array}
\end{equation}
where $A=(a_{ij})_{i,j=0,1,2}$ and $U=(u_{ij})_{i,j=0,1,2}$ are given by
eq.(\ref{AandU}).  Then we can verify the following:
\begin{prop}
 The transformations $s_0$, $s_1$, $s_2$ and $\pi$ 
of the $A$-variables and the $F$-variables, 
defined by eq.$($\ref{ultraW}$)$
generate the extended affine Weyl group 
$\widetilde{W}=\br{s_0,s_1,s_2,\pi}$ of type $A^{(1)}_2$. 
Furthermore, they commute with the time evolution $t$ 
of the fourth ultra-discrete Painlev\'e equation uP$_{\IV}$. 
\end{prop}

%%%%%%%%%%%%%%%%%%%%%%%%%%%%%%%%%%%%%%%%%%%%%%%%%%%%%%%%
\appendix
\section{Difference Formulas and Pl\"ucker Relations}
In this appendix, we provide with data which are necessary for the proof of
Proposition \ref{bilinear}.

We first note that it is possible to express $\phi_N^{\nu,m}$ in Section
2.2 as Casorati determinant with respect to $m$ as follows.
\begin{lem}\label{various phi}
$\phi_N^{\nu,m}$ is rewritten as
\begin{eqnarray}
&&  \phi_N^{\nu,m}=\left|
\begin{array}{cccc}
 G_\nu^{m}    &G_{\nu-1}^{m}   &\cdots & G_{\nu-N+1}^{m}\\
 G_\nu^{m+1}  &G_{\nu-1}^{m+1} &\cdots &G_{\nu-N+1}^{m+1} \\
 \vdots & \vdots & \ddots & \vdots\\
 G_\nu^{m+N-1}&G_{\nu-1}^{m+N-1} &\cdots &G_{\nu-N+1}^{m+N-1} 
\end{array}
\right|\ ,\label{phi:2} \\
&&  \phi_N^{\nu,m}=\prod_{k=1}^{N-1}\left(\frac{a_0^2q^{2(n-k)}}
{a_0^2q^{2(n-k)}-1}\right)^{N-k}~\left|
\begin{array}{cccc}
 G_\nu^{m}&G_\nu^{m+1} &\cdots &G_\nu^{m+N-1} \\
 G_{\nu}^{m-1}&G_{\nu}^{m} &\cdots &G_{\nu}^{m+N-2} \\
 \vdots & \vdots & \ddots & \vdots\\
 G_{\nu}^{m-N+1}&G_{\nu}^{m-N+2} &\cdots &G_{\nu}^{m} 
\end{array}
\right|\ .\label{phi:3}
\end{eqnarray}
\end{lem}
\noindent{\it Proof of Lemma \ref{various phi}.}
We use the contiguity relation (\ref{contiguity:cn2}), which is rewritten
with the notation introduced in eq.(\ref{Gnm}) as
\begin{equation}
 G_\nu^{m+1} = G_{\nu+1}^m - q^{2m}c^2 G_\nu^m.\label{qHermite:c3}
\end{equation}
For $k=2$ to $N$, subtracting $(j-1)$-st column multiplied by $q^{2m}c^2$ from $j$-th
column of eq.(\ref{phi:1}) for $j=N,\cdots k$, we obtain 
\begin{eqnarray}
 \phi_N^{\nu,m}&=&\left|
\begin{array}{cccc}
 G_\nu^m    &G_{\nu+1}^{m}-q^{2m}c^2 G_\nu^m&\cdots &G_{\nu+N-1}^m-q^{2m}c^2G_{\nu+N-2}^m \\
 G_{\nu-1}^m&G_{\nu}^m-q^{2m}c^2 G_{\nu-1}^m &\cdots &G_{\nu-N-2}^m-q^{2m}c^2G_{\nu+N-3}^m \\
\vdots & \vdots & \ddots & \vdots\\
 G_{\nu-N+1}^m&G_{\nu-N+2}^m-q^{2m}c^2G_{\nu-N+1}^m &\cdots &G_{\nu}^m-q^{2m}c^2G_{\nu-1}^m 
\end{array}\right|\nonumber \\
&=&\left|
\begin{array}{cccc}
 G_\nu^m      &G_{\nu}^{m+1}     &\cdots &G_{\nu+N-2}^{m+1}\\
 G_{\nu-1}^m  &G_{\nu-1}^{m+1}   &\cdots &G_{\nu+N-3}^{m+1} \\
\vdots & \vdots & \ddots & \vdots\\
 G_{\nu-N+1}^m&G_{\nu-N+1}^{m+1} &\cdots &G_{\nu-1}^{m+1} 
\end{array}\right|\nonumber\\
&=&\cdots\nonumber\\
&=&\left|
\begin{array}{cccc}
 G_\nu^{m}    &G_{\nu}^{m+1}   &\cdots & G_{\nu}^{m+N-1}\\
 G_{\nu-1}^{m}  &G_{\nu-1}^{m+1} &\cdots &G_{\nu-1}^{m+N-1} \\
 \vdots & \vdots & \ddots & \vdots\\
 G_{\nu-N+1}^{m}&G_{\nu-N+1}^{m+1} &\cdots &G_{\nu-N+1}^{m+N-1} 
\end{array}
\right|,\label{casorati:1}
\end{eqnarray}
which yields eq.(\ref{phi:2}) by taking the transposition.
In order to derive eq.(\ref{phi:3}), we use the contiguity relation,
\begin{equation}
 G_{\nu-1}^m + \frac{G_\nu^m}{a_0^2q^{2(\nu-1)}-1} = \frac{a_0^2q^{2(\nu-1)}}{a_0^2q^{2(\nu-1)}-1}~G_{\nu}^{m-1},
\end{equation}
which is obtained from eqs.(\ref{contiguity:cn1}) and (\ref{contiguity:cn2}).
For $k=2$ to $N$, subtracting $(i-1)$-st row multiplied by $\frac{1}{a_0^2q^{2(\nu-i-1)}-1}$ from $i$-th
row of eq.(\ref{casorati:1}) for $i=N,\cdots k$, we obtain 
\begin{eqnarray*}
\phi_N^{\nu,m} &=&\left|
\begin{array}{cccc}
 G_\nu^{m}    &G_{\nu}^{m+1}   &\cdots & G_{\nu}^{m+N-1}\\
 \vdots & \vdots & \ddots & \vdots\\
 G_{\nu-N+2}^{m}  &G_{\nu-N+2}^{m+1} &\cdots &G_{\nu-N+2}^{m+N-1} \\
 G_{\nu-N+1}^{m}-\frac{G_{\nu-N+2}^m}{a_0^2q^{2(\nu-N+1)}-1}&G_{\nu-N+1}^{m+1}-\frac{G_{\nu-N+2}^{m+1}}
{a_0^2q^{2(\nu-N+1)}-1} &\cdots &G_{\nu-N+1}^{m+N-1}-\frac{G_{\nu-N+2}^{m+N-1}}{a_0^2q^{2(\nu-N+1)}-1} 
\end{array}
\right|\\
&=&\frac{a_0^2q^{2(\nu-N+1)}}{a_0^2q^{2(\nu-N+1)}-1}\left|
\begin{array}{cccc}
 G_\nu^{m}    &G_{\nu}^{m+1}   &\cdots & G_{\nu}^{m+N-1}\\
 \vdots & \vdots & \ddots & \vdots\\
 G_{\nu-N+2}^{m}  &G_{\nu-N+2}^{m+1} &\cdots &G_{\nu-N+2}^{m+N-1} \\
G_{\nu-N+2}^{m-1}&G_{\nu-N+2}^{m}&\cdots &G_{\nu-N+2}^{m+N-2}
\end{array}
\right|\\
&=&\cdots\\
&=&\prod_{i=1}^{N-1}\frac{a_0^2q^{2(\nu-i)}}{a_0^2q^{2(\nu-i)}-1}\left|
\begin{array}{cccc}
 G_\nu^{m}    &G_{\nu}^{m+1}   &\cdots & G_{\nu}^{m+N-1}\\
 G_{\nu}^{m-1}  &G_{\nu}^{m} &\cdots &G_{\nu}^{m+N-2} \\
 \vdots & \vdots & \ddots & \vdots\\
 G_{\nu-N+2}^{m}  &G_{\nu-N+2}^{m+1} &\cdots &G_{\nu-N+2}^{m+N-1} \\
G_{\nu-N+2}^{m-1}&G_{\nu-N+2}^{m}&\cdots &G_{\nu-N+2}^{m+N-2}
\end{array}
\right|\\
&=&\cdots\\
&=&\prod_{i=1}^{N-1}\left(\frac{a_0^2q^{2(\nu-i)}}{a_0^2q^{2(\nu-i)}-1}\right)^{N-i}\left|
\begin{array}{cccc}
 G_\nu^{m}    &G_{\nu}^{m+1}   &\cdots & G_{\nu}^{m+N-1}\\
 G_{\nu}^{m-1}  &G_{\nu}^{m} &\cdots &G_{\nu}^{m+N-2} \\
 \vdots & \vdots & \ddots & \vdots\\
 G_{\nu}^{m-N+2}  &G_{\nu}^{m-N+3} &\cdots &G_{\nu}^{m-1} \\
G_{\nu}^{m-N+1}&G_{\nu}^{m-N}&\cdots &G_{\nu}^{m}
\end{array}
\right|,
\end{eqnarray*}
which is eq.(\ref{phi:3}).\hfill\qed

In view of this lemma, we introduce the following notations.
\begin{eqnarray}
 \phi_N^{\nu,m}&=&
\left|
\begin{array}{cccc}
 G_\nu^m&G_{\nu+1}^m &\cdots &G_{\nu+N-1}^m \\
 G_{\nu-1}^m&G_{\nu}^m &\cdots &G_{\nu-N-2}^m \\
\vdots & \vdots & \ddots & \vdots\\
 G_{\nu-N+1}^m&G_{\nu-N+2}^m &\cdots &G_{\nu}^m
\end{array}\right| 
=\left|\mathbf{0}^m,\ \mathbf{1}^m,\ \cdots,\ \mathbf{N-1}^m\right|,\label{phi0}\\
 \phi_N^{\nu,m}&=&
\left|
\begin{array}{cccc}
 G_\nu^{m}    &G_{\nu-1}^{m}   &\cdots & G_{\nu-N+1}^{m}\\
 G_\nu^{m+1}  &G_{\nu-1}^{m+1} &\cdots &G_{\nu-N+1}^{m+1} \\
 \vdots & \vdots & \ddots & \vdots\\
 G_\nu^{m+N-1}&G_{\nu-1}^{m+N-1} &\cdots &G_{\nu-N+1}^{m+N-1} 
\end{array}
\right|=
|\underline{\mathbf{0}}^{m-1},\ \underline{\mathbf{-1}}^{m-1},
\cdots,\underline{\mathbf{-N+1}}^{m-1}|,\label{phi1}\\
 \psi_N^{\nu,m}&=&\left|
\begin{array}{cccc}
 G_\nu^{m}&G_\nu^{m+1} &\cdots &G_\nu^{m+N-1} \\
 G_{\nu}^{m-1}&G_{\nu}^{m} &\cdots &G_{\nu}^{m+N-2} \\
 \vdots & \vdots & \ddots & \vdots\\
 G_{\nu}^{m-N+1}&G_{\nu}^{m-N+2} &\cdots &G_{\nu}^{m} 
\end{array}
\right| = |\overline{\mathbf{0}}_\nu, \ \overline{\mathbf{1}}_\nu, \cdots,
\overline{\mathbf{N-2}}_\nu,\ \overline{\mathbf{N-1}}_\nu |, \label{psi}
\end{eqnarray}
where $\mathbf{k}^m$, $\underline{\mathbf{k}}^m$ and $\overline{\mathbf{k}}_\nu$ are
column vectors given by
\begin{equation}
 \mathbf{k}^m=\left(\begin{array}{c}G_{\nu+k}^m \\G_{\nu+k-1}^m\\\vdots\\G_{\nu+k-N+1}^m
		  \end{array}\right),\quad
 \underline{\mathbf{k}}^{m}=\left(\begin{array}{c}G_{\nu+k}^m  \\
G_{\nu+k}^{m+1}\\\vdots\\G_{\nu+k}^{m+N-1}
  \end{array}\right),\quad  \overline{\mathbf{k}}_\nu 
= \left(\begin{array}{c}G_\nu^{m+k} \\G_\nu^{m+k-1}\\\vdots\\
G_\nu^{m+k-N+1} \end{array}\right).
\end{equation}
respectively. We also use auxiliary column vectors,
\begin{equation}
\mathbf{k}^{m\prime}=
\left(\begin{array}{c}G_{\nu+k}^m \\q^2G_{\nu+k-1}^m\\\vdots\\q^{2(N-1)}G_{\nu+k-N+1}^m
\end{array}\right),\quad
 \underline{\mathbf{k}}^{m\dag}=\left(\begin{array}{c}G_{\nu+k}^m  \\q^{-2}G_{\nu+k}^{m+1}\\\vdots\\
q^{-2(N-1)}G_{\nu+k}^{m+N-1}  \end{array}\right),\quad
\overline{\mathbf{k}}_\nu'= \left(\begin{array}{c}G_\nu^{m+k} \\q^2G_\nu^{m+k-1}\\\vdots\\
q^{2(N-1)}G_\nu^{m+k-N+1} \end{array}\right).
\end{equation}

Now we give difference formulas.
\par\medskip
\noindent{\bf Difference Formula I}
\begin{equation}
\begin{array}{l}
\medskip
 |\mathbf{0}^{m+1\prime},\ \mathbf{0}^m,\ \mathbf{1}^m,\cdots,\ \mathbf{N-3}^m,\
 \mathbf{N-2}^m| = a_0^{-2(N-1)}c^{-2(N-1)}q^{-2(N-1)(m+\nu-1)}~\phi_N^{\nu,m+1},\\
 |\mathbf{1}^{m+1\prime},\ \mathbf{0}^m,\ \mathbf{1}^m,\cdots,\ \mathbf{N-3}^m,\
 \mathbf{N-2}^m| = a_0^{-2(N-1)}c^{-2(N-1)}q^{-2(N-1)(m+\nu-1)}~\phi_N^{\nu,m+1}.
\end{array}
 \label{Difference Formula I}
\end{equation}

\noindent{\bf Difference Formula II}
\begin{equation}
 \begin{array}{l}
\medskip
   |\mathbf{0}^{m},\ \mathbf{0}^{m+1},\ \mathbf{1}^{m+1},\cdots,\ \mathbf{N-3}^{m+1},\
 \mathbf{N-2}^{m+1}| = \phi_N^{\nu,m},\\
|\mathbf{1}^{m},\ \mathbf{0}^{m+1},\ \mathbf{1}^{m+1},\cdots,\ \mathbf{N-3}^{m+1},\
 \mathbf{N-2}^{m+1}| = c^2q^{2m}~\phi_N^{\nu,m}.
 \end{array}
\label{Difference Formula II}
\end{equation}

\noindent{\bf Difference formula III}
\begin{equation}
 \begin{array}{l}
\medskip
|\underline{\mathbf{-1}}^{m-1},\ \underline{\mathbf{-2}}^{m-1},\cdots,\underline{\mathbf{-N+1}}^{m-1},
\ \underline{\mathbf{-N+1}}^{m\dag}|
=a_0^{-2(N-1)}c^{-2(N-1)}q^{-2(N-1)(\nu+m-1)}~\phi_N^{\nu,m},  \\
|\underline{\mathbf{-1}}^{m-1},\ \underline{\mathbf{-2}}^{m-1},
\cdots,\underline{\mathbf{-N+1}}^{m-1},\ \underline{\mathbf{-N+2}}^{m\dag}|
=a_0^{-2(N-1)}c^{-2(N-1)}q^{-2(N-1)(\nu+m-1)}~\phi_N^{\nu,m}.
 \end{array}
\label{Difference Formula III}
\end{equation}

\noindent{\bf Difference Formula IV}
\begin{equation}
\begin{array}{l}
\medskip
|\overline{\mathbf{0}}_{\nu+1}',\ \overline{\mathbf{0}}_{\nu},\cdots,\overline{\mathbf{N-2}}_{\nu}|
=c^{-2(N-1)}(a_0^2q^{2\nu}-1)^{-(N-1)}q^{-2(m-1)(N-1)}~\psi_N^{\nu+1,m},\\ 
|\overline{\mathbf{1}}_{\nu+1}',\ \overline{\mathbf{0}}_{\nu},\cdots,\overline{\mathbf{N-2}}_{\nu}|
=c^{-2(N-1)}(a_0^2q^{2\nu}-1)^{-(N-1)}q^{-2(m-1)(N-1)}~\psi_N^{\nu+1,m}.
\end{array}
\label{Difference Formula IV}
\end{equation}

\noindent{\bf Difference Formula V}
\begin{equation}
 \begin{array}{l}
\medskip
  |\overline{\mathbf{0}}_{\nu+1},\ \overline{\mathbf{0}}_{\nu},\cdots,\overline{\mathbf{N-2}}_{\nu}|
=(1-a_0^2q^{2\nu})^{-(N-1)}~\psi_N^{\nu+1,m},\\
|\overline{\mathbf{1}}_{\nu+1},\ \overline{\mathbf{0}}_{\nu},\cdots,\overline{\mathbf{N-2}}_{\nu}|
=(1-a_0^2q^{2\nu})^{-(N-1)}a_0^2q^{2\nu}~\psi_N^{\nu+1,m}.
 \end{array}
\label{Difference Formula V}
\end{equation}
Difference formula I has been proved in Lemma \ref{Difference Formula
I:lem}. Other formulas are proved in a similar manner.
Table \ref{table1} shows the the contiguity relation and determinant
to be used for the derivation of each difference formula.
\begin{table}[h]
\renewcommand{\arraystretch}{1.5}
\caption{Data for the proof of difference formulas\label{table1}}
\begin{center}
\begin{tabular}{|c||c|c|}
\hline
 \textbf{Difference Formula}&\textbf{Contiguity Relation}
 &\textbf{Determinant} \\
\hline
Difference Formula I (\ref{Difference Formula I})
 & $ G_{\nu+1}^{m+1}=G_{\nu}^{m+1}+a_0^2q^{2(\nu+m)}c^2G_\nu^m$ 
&eq.(\ref{phi0})\\
\hline
Difference Formula II(\ref{Difference Formula II}) & $ G_{\nu}^{m+1}=G_{\nu+1}^m - q^{2m}c^2 G_\nu^m$ 
&eq.(\ref{phi0})\\
\hline
Difference Formula III(\ref{Difference Formula III}) & $ G_{\nu}^{m}-G_{\nu-1}^{m} = a_0^2c^2 q^{2(\nu+m-2)}G_{\nu-1}^{m-1}$
&eq.(\ref{phi1})\\
\hline
Difference Formula IV(\ref{Difference Formula IV}) & $G_{\nu+1}^{m+1}-G_{\nu+1}^{m} = c^2 q^{2m}(1-a_0q^{2\nu})G_{\nu}^{m}$
&eq.(\ref{psi})\\
\hline
Difference Formula V(\ref{Difference Formula V}) &
 $G_{\nu+1}^{m+1}-a_0q^{2\nu}G_{\nu+1}^{m} = (1-a_0q^{2\nu})G_{\nu}^{m+1}$
&eq.(\ref{psi})\\
\hline
\end{tabular}
\end{center}
\end{table}

Next we give the list of Pl\"ucker relations which are necessary for the
proof of Proposition \ref{bilinear}. 

\noindent\textbf{Pl\"ucker Relation I}
\begin{eqnarray}
  0&=&\left|\varphi_1,\ \mathbf{1}^{m+1\prime},\ \mathbf{1}^{m},\cdots,\mathbf{N-1}^{m}\right|
\times\left|\mathbf{0}^{m},\ \mathbf{1}^{m},\cdots,\mathbf{N-1}^{m},\ \mathbf{N}^{m}\right|\nonumber\\
   &-&\left|\varphi_1,\ \mathbf{0}^{m},\ \mathbf{1}^{m},\cdots,\mathbf{N-1}^{m}\right|
\times\left|\mathbf{1}^{m+1\prime},\ \mathbf{1}^{m},\cdots,\mathbf{N-1}^{m},\ \mathbf{N}^{m}\right|\nonumber\\
   &+&\left|\mathbf{1}^{m+1\prime},\ \mathbf{0}^{m},\ \mathbf{1}^{m},\cdots,\mathbf{N-1}^{m}\right|
\times\left|\varphi_1,\ \mathbf{1}^{m},\cdots,\mathbf{N-1}^{m},\ \mathbf{N}^{m}\right|.\label{pl1}
\end{eqnarray}
\noindent\textbf{Pl\"ucker Relation II}
\begin{eqnarray}
  0&=&\left|\varphi_2,\ \mathbf{1}^{m+1\prime},\ \mathbf{1}^{m},\cdots,\mathbf{N-1}^{m}\right|
\times\left|\mathbf{0}^{m},\ \mathbf{1}^{m},\cdots,\mathbf{N-1}^{m},\ \mathbf{N}^{m}\right|\nonumber\\
   &-&\left|\varphi_2,\ \mathbf{0}^{m},\ \mathbf{1}^{m},\cdots,\mathbf{N-1}^{m}\right|
\times\left|\mathbf{1}^{m+1\prime},\ \mathbf{1}^{m},\cdots,\mathbf{N-1}^{m},\ \mathbf{N}^{m}\right|\nonumber\\
   &+&\left|\mathbf{1}^{m+1\prime},\ \mathbf{0}^{m},\ \mathbf{1}^{m},\cdots,\mathbf{N-1}^{m}\right|
\times\left|\varphi_2,\ \mathbf{1}^{m},\cdots,\mathbf{N-1}^{m},\ \mathbf{N}^{m}\right|.\label{pl2}
\end{eqnarray}
\noindent\textbf{Pl\"ucker Relation III}
\begin{eqnarray}
0&=&\left|\varphi_2,\ \mathbf{1}^{m},\ \mathbf{1}^{m+1},\cdots,\mathbf{N-1}^{m+1}\right|
\times\left|\mathbf{0}^{m+1},\ \mathbf{1}^{m+1},\cdots,\mathbf{N-1}^{m+1},\ \mathbf{N}^{m+1}\right|
\nonumber\\
&-&\left|\varphi_2,\ \mathbf{0}^{m+1},\ \mathbf{1}^{m+1},\cdots,\mathbf{N-1}^{m+1}\right|
\times\left|\mathbf{1}^{m},\ \mathbf{1}^{m+1},\cdots,\mathbf{N-1}^{m+1},\ \mathbf{N}^{m+1}\right|\nonumber\\
   &+&\left|\mathbf{1}^{m},\ \mathbf{0}^{m+1},\ \mathbf{1}^{m+1},\cdots,\mathbf{N-1}^{m+1}\right|
\times\left|\varphi_2,\ \mathbf{1}^{m+1},\cdots,\mathbf{N-1}^{m+1},\ \mathbf{N}^{m+1}\right| .\label{pl3}
\end{eqnarray}
\noindent\textbf{Pl\"ucker Relation IV}
\begin{eqnarray}
0&=&\left|\varphi_2,\ \varphi_1,\ \mathbf{1}^{m+1},\cdots,\mathbf{N-1}^{m+1}\right|
\times\left|\mathbf{1}^{m},\ \mathbf{1}^{m+1},\cdots,\mathbf{N-1}^{m+1},\ \mathbf{N}^{m+1}\right|\nonumber\\
   &-&\left|\varphi_2,\ \mathbf{1}^{m},\ \mathbf{1}^{m+1},\cdots,\mathbf{N-1}^{m+1}\right|
\times\left|\varphi_1,\ \mathbf{1}^{m+1},\cdots,\mathbf{N-1}^{m+1},\ \mathbf{N}^{m+1}\right|\nonumber\\
   &+&\left|\varphi_1,\ \mathbf{1}^{m},\ \mathbf{1}^{m+1},\cdots,\mathbf{N-1}^{m+1}\right|
\times\left|\varphi_2,\ \mathbf{1}^{m+1},\cdots,\mathbf{N-1}^{m+1},\ \mathbf{N}^{m+1}\right| .\label{pl4}
\end{eqnarray}
\noindent\textbf{Pl\"ucker Relation V}
\begin{eqnarray}
0&=&\left|\varphi_2,\ \varphi_1,\ \mathbf{0}^{m},\cdots,\mathbf{N-2}^{m}\right|
\times\left|\mathbf{0}^{m+1\prime},\ \mathbf{0}^{m},\cdots,\mathbf{N-2}^{m},\ \mathbf{N-1}^{m}\right|\nonumber\\
   &-&\left|\varphi_2,\ \mathbf{0}^{m+1\prime},\ \mathbf{0}^{m},\cdots,\mathbf{N-2}^{m}\right|
\times\left|\varphi_1,\ \mathbf{0}^{m},\cdots,\mathbf{N-2}^{m},\ \mathbf{N-1}^{m}\right|\nonumber\\
   &+&\left|\varphi_1,\ \mathbf{0}^{m+1\prime},\ \mathbf{0}^{m},\cdots,\mathbf{N-2}^{m}\right|
\times\left|\varphi_2,\ \mathbf{0}^{m},\cdots,\mathbf{N-2}^{m},\ \mathbf{N-1}^{m}\right| .\label{pl4aux}
\end{eqnarray}
\noindent\textbf{Pl\"ucker Relation VI}
\begin{eqnarray}
 0&=&|\underline{\mathbf{0}}^{m-1},\ \underline{\mathbf{-1}}^{m-1},\cdots,
\underline{\mathbf{-N+1}}^{m-1},\ \underline{\mathbf{-N}}^{m-1}|\times
|\underline{\mathbf{-1}}^{m-1},\cdots,\underline{\mathbf{-N+1}}^{m-1},
\ \underline{\mathbf{-N+1}}^{m+1\dag},\ \varphi_1|\nonumber\\
&-&|\underline{\mathbf{0}}^{m-1},\ \underline{\mathbf{-1}}^{m-1},\cdots,
\underline{\mathbf{-N+1}}^{m-1},\ \underline{\mathbf{-N+1}}^{m\dag}|\times
|\underline{\mathbf{-1}}^{m-1},\cdots,\underline{\mathbf{-N+1}}^{m-1},
\ \underline{\mathbf{-N}}^{m-1},\ \varphi_1|\nonumber\\ 
&+&|\underline{\mathbf{0}}^{m-1},\ \underline{\mathbf{-1}}^{m-1},\cdots,
\underline{\mathbf{-N+1}}^{m-1},\ \varphi_1|\times
|\underline{\mathbf{-1}}^{m-1},\cdots,\underline{\mathbf{-N+1}}^{m-1},\ \underline{\mathbf{-N}}^{m-1},
\ \underline{\mathbf{-N+1}}^{m\dag}|.\label{pl5}
\end{eqnarray}
\noindent\textbf{Pl\"ucker Relation VII}
\begin{eqnarray}
0&=&\left|\varphi_1,\ \overline{\mathbf{1}}_{\nu+1}',\ \overline{\mathbf{1}}_{\nu},\cdots
,\overline{\mathbf{N-1}}_{\nu}\right|
\times\left|\overline{\mathbf{0}}_{\nu},\ \overline{\mathbf{1}}_{\nu},\cdots
,\overline{\mathbf{N-1}}_{\nu},\ \overline{\mathbf{N}}_{\nu}\right|\nonumber\\
   &-&\left|\varphi_1,\ \overline{\mathbf{0}}_{\nu},\ \overline{\mathbf{1}}_{\nu}
,\cdots,\overline{\mathbf{N-1}}_{\nu}\right|
\times\left|\overline{\mathbf{1}}_{\nu+1}',\ \overline{\mathbf{1}}_{\nu}
,\cdots,\overline{\mathbf{N-1}}_{\nu},\ \overline{\mathbf{N}}_{\nu}\right|\nonumber\\
   &+&\left|\overline{\mathbf{1}}_{\nu+1}',\ \overline{\mathbf{0}}_{\nu}
,\ \overline{\mathbf{1}}_{\nu},\cdots,\overline{\mathbf{N-1}}_{\nu}\right|
\times\left|\varphi_1,\ \overline{\mathbf{1}}_{\nu},\cdots,\overline{\mathbf{N-1}}_{\nu},\ 
\overline{\mathbf{N}}_{\nu}\right|. \label{pl6}
\end{eqnarray}
\noindent\textbf{Pl\"ucker Relation VIII}
\begin{eqnarray}
 0&=&|\varphi_2,\ \varphi_1, \overline{\mathbf{1}}_{\nu},\cdots,
\overline{\mathbf{N-1}}_{\nu}|\times
|\overline{\mathbf{0}}_{\nu+1},\ \overline{\mathbf{1}}_{\nu},\cdots,\overline{\mathbf{N-1}}_{\nu},
\ \overline{\mathbf{N}}_{\nu}|\nonumber\\
&-&|\varphi_2,\ \overline{\mathbf{0}}_{\nu+1},\ \overline{\mathbf{1}}_{\nu},\cdots,\overline{\mathbf{N-1}}_{\nu}|\times
|\varphi_1,\ \overline{\mathbf{1}}_{\nu},\cdots,\overline{\mathbf{N-1}}_{\nu},\overline{\mathbf{N}}_{\nu}|\nonumber\\
&+&|\varphi_1,\ \overline{\mathbf{0}}_{\nu+1},\ \overline{\mathbf{1}}_{\nu},\cdots,\overline{\mathbf{N-1}}_{\nu}|\times
|\varphi_2,\ \overline{\mathbf{1}}_{\nu},\cdots,\overline{\mathbf{N-1}}_{\nu},\overline{\mathbf{N}}_{\nu}|.\label{pl7}
\end{eqnarray}

\begin{table}[h]
 \begin{center}
\caption{Data for the proof of bilinear equations\label{table2}}
\renewcommand{\arraystretch}{1.5}
  \begin{tabular}{|c||l|l|}
\hline
  \textbf{Bilinear Equation}&\textbf{Pl\"ucker Relation}
   &\textbf{Difference Formula} \\
\hline
eq.(\ref{bl1}) & Pl\"ucker Relation I (\ref{pl1}) & Difference Formula I (\ref{Difference
   Formula I})\\
\hline
   eqs.(\ref{bl2}) and (\ref{bl2:aux})& \hskip-6pt
\begin{tabular}{l}
Pl\"ucker Relation II (\ref{pl2})\\
Pl\"ucker Relation III (\ref{pl3})  
\end{tabular}&\hskip-6pt
\begin{tabular}{l}
Difference Formula I (\ref{Difference Formula I}) \\ 
Difference Formula II (\ref{Difference Formula II})
\end{tabular}\\
\hline
eq.(\ref{bl3})& Pl\"ucker Relation IV(\ref{pl4}) & Difference Formula II (\ref{Difference
   Formula II}) \\
\hline
eq.(\ref{bl3:aux})& Pl\"ucker Relation V(\ref{pl4aux}) & Difference Formula I (\ref{Difference
   Formula I}) \\
\hline
eq.(\ref{bl4})&  Pl\"ucker Relation VI(\ref{pl5}) & Difference Formula III (\ref{Difference
   Formula III}) \\
\hline
eq.(\ref{bl5})&  Pl\"ucker Relation VII(\ref{pl6}) & Difference Formula IV (\ref{Difference
   Formula IV}) \\
\hline
eq.(\ref{bl6})& Pl\"ucker Relation VIII(\ref{pl7}) & Difference Formula V (\ref{Difference
   Formula V})\\
\hline
  \end{tabular}
 \end{center}
\end{table}
Here, $\varphi_i$ $(i=1,2)$ are any column vectors which we specialize as
\begin{equation}
\varphi_1=\left(\begin{array}{c}1\\0\\ \vdots\\0 \end{array}\right),\quad
 \varphi_2=\left(\begin{array}{c}0\\\vdots\\0 \\1 \end{array}\right).
\end{equation}
in the proof of bilinear equations.

In Section 2.2.2, we have shown how eq.(\ref{bl1}) can be derived from
the Pl\"ucker relation I and difference formula I. The other bilinear
equations in Proposition \ref{bilinear} are obtained similarly by using
the data described in Table \ref{table2}.

%%%%%%%%%%%%%%%%%%%%%%%%%%%%%%%%%%%%%%%%%%%%%%%%%%%%%%%%%%%%%%%%%%%%%%%%%%%%%%%%%%%%%%%%%%%%

\end{document}